\documentclass[10pt,twocolumn]{article}
\usepackage[top=1.8cm, bottom=1.8cm, left=1.8cm, right=1.8cm]{geometry}

\usepackage{times}
\renewcommand{\footnotesize}{\fontsize{8}{9}\selectfont}
\usepackage[compact]{titlesec}
\titlespacing*{\section}{0pt}{*3}{3pt}
\titlespacing{\subsection}{0pt}{*2}{2pt}
\titlespacing{\subsubsection}{0pt}{*2}{2pt}

\usepackage[hang,flushmargin]{footmisc}

\renewcommand{\footnoterule}{%
  \kern -3pt
  \hrule width 1in
  \kern 2pt
}

\usepackage{url}
\makeatletter
\def\url@leostyle{%
  \@ifundefined{selectfont}{\def\UrlFont{}}%
  {\def\UrlFont{}}%
}
\makeatother
\usepackage[hyphenbreaks]{breakurl}

\usepackage{graphicx}  %
\usepackage{multirow}
\usepackage{booktabs}
\usepackage{subfigure}
\usepackage{fdsymbol}
\usepackage[compact]{titlesec}
\usepackage{xcolor}
\usepackage{colortbl}
\usepackage[keeplastbox]{flushend}

\graphicspath{ {figures/} }

\renewenvironment{thebibliography}[1]{
  \begin{oldthebibliography}{#1}
    \setlength{\itemsep}{0.0em}
    \setlength{\parskip}{0.0em}
}
{
  \end{oldthebibliography}
}

\usepackage{paralist}
\newcommand{\reduce}{\vspace{-2pt}}

\newif\ifcomment
\commentfalse
\ifcomment
\newcommand{\yw}[1]{{\bf \textcolor{purple}{YW: #1}}}
\newcommand{\sz}[1]{{\bf \textcolor{brown}{SZ: #1}}}
\newcommand{\jbnote}[1]{{\bf \textcolor{magenta}{JB: #1}}}
\definecolor{ftGreen}{rgb}{0.0, 0.5, 0.0}
\newcommand{\ft}[1]{{\bf \textcolor{ftGreen}{FT: #1}}}
\newcommand{\edc}[1]{{\bf\textcolor{blue}{EDC: #1}}}
\newcommand{\gs}[1]{{\bf\textcolor{red}{GS: #1}}}
\newcommand{\bb}[1]{{\bf\textcolor{orange}{BB: #1}}}
\newcommand{\dm}[1]{{\bf\textcolor{grey}{DM: #1}}}
\else
\newcommand{\yw}[1]{}
\newcommand{\sz}[1]{}
\newcommand{\ft}[1]{}
\newcommand{\jbnote}[1]{}
\newcommand{\edc}[1]{}
\newcommand{\gs}[1]{}
\newcommand{\bb}[1]{}
\newcommand{\dm}[1]{}
\fi

\usepackage[small,bf]{caption}

\definecolor{linkcol}{RGB}{0,70,25}
\definecolor{citecol}{RGB}{0,70,25}
\definecolor{urlcol}{RGB}{0,70,25}
\usepackage[bookmarks=true,bookmarksnumbered=true,colorlinks=true,linkcolor=linkcol,citecolor=citecol,urlcolor=urlcol,hypertexnames=true,pdfpagelabels]{hyperref}

\newcommand{\descr}[1]{\smallskip\noindent\textbf{#1}}

\title{\bf Understanding the Use of Fauxtography on Social Media}
\author{Yuping Wang$^{\clubsuit}$, Fatemeh Tahmasbi$^{\vardiamondsuit}$, Jeremy Blackburn$^{\vardiamondsuit}$, Barry Bradlyn$^{\spadesuit}$,\\Emiliano De Cristofaro$^\varheartsuit$, David Magerman$^\partial$, Savvas Zannettou$^\bigstar$, and Gianluca Stringhini$^{\clubsuit}$\\[0.5ex]
\normalsize $^{\clubsuit}$Boston University, $^{\vardiamondsuit}$Binghamton University, $^\spadesuit$University of Illinois at Urbana--Champaign, \\ \normalsize $^\varheartsuit$University College London, $^\partial$  Differential Venture Partners, $^\bigstar$Max Planck Institute for Informatics}

\begin{document}
\date{}

\maketitle

\begin{abstract}
Despite the influence that image-based communication has on online discourse, the role played by images in disinformation is still not well understood.
In this paper, we present the first large-scale study of \emph{fauxtography}, analyzing the use of manipulated or misleading images in news discussion on online communities.
First, we develop a computational pipeline geared to detect fauxtography, and identify over 61k instances of fauxtography discussed on Twitter, 4chan, and Reddit.
Then, we study how posting fauxtography affects engagement of posts on social media, finding that posts containing it receive more interactions in the form of re-shares, likes, and comments.
Finally, we show that fauxtography images are often turned into memes by Web communities.
Our findings show that effective mitigation against disinformation need to take images into account, and highlight a number of challenges in dealing with image-based disinformation.
\end{abstract}

\section{Introduction}
Recent years have seen an increase in false information published online and spread through social media~\cite{kumar2018false}.
An important aspect of news consumption is that users not only pay attention to text, but also to the accompanying images in the article.
In fact, research in psychology shows that images play a crucial role in both how readers perceive certain issues~\cite{zillmann1999effects} and in which articles individuals choose to read~\cite{zillmann2001effects}.
Therefore, it is not surprising that images may be manipulated or misrepresented to mislead users.

In this paper, we focus on {\em fauxtography}~\cite{cooper2007concise}, i.e., news images that have been modified or miscaptioned to change their intent, often with the goal of spreading a false sense of the events they purport to depict.
Although previous research efforts have proposed detection tools for fauxtography~\cite{zhang2018fauxbuster,zlatkova2019fact}, to the best our knowledge, the {\em impact} of fauxtography on news discussion has not been studied.
In particular, we set out to investigate two research questions:
\begin{itemize}
\item RQ1: Does sharing fauxtography increase engagement on social media?
\item RQ2: Do fauxtography images have a life beyond their questionable verisimilitude (their appearance of being being real)? I.e., do new variants and memes using them appear on social media?
    \end{itemize}

To answer these questions, we develop a computational analysis pipeline geared to identify posts containing fauxtography at scale, measure the engagement of users sharing and viewing such posts, and understand how these images are used on different social media platforms.
First, we gather 2.6 billion posts from three social media platforms (Twitter, Reddit, and 4chan) as well as 32M news articles published by over 1,000 news websites.
Then, we extract all images appearing in these posts and articles, and use perceptual hashing~\cite{monga2006perceptual} to match them to images labeled as fauxtography by the fact-checking site Snopes.
In total, we identify 61K posts containing fauxtography shared by users over the two year period from 2016 to 2018.

To address RQ1, we analyze the reactions to posts containing fauxtography on social media, compared with the reaction to posts by the same users with no image or with images characterized as non-fauxtography.
We find that including fauxtography in posts does increase user engagement on social media. %
On the other hand, posting links to news articles that contain fauxtography (rather than posting images directly) does not increase engagement on Twitter, while it does yield more interactions on Reddit.
Surprisingly, the extent to which an image is misleading -- e.g., whether it is completely false or just partially true -- does not significantly affect engagement, suggesting that the increased engagement is driven by the inflammatory and controversial nature of fauxtography more than its verisimilitude.

For RQ2, we search for variants of fauxtography images that each appear on all of the social media platforms.
Our intiution is that instances of fauxtography are likely have some sort of inherent exploitability making them suitable as a base for new memes. Visual memes have become important to the spread of racist and political ideology~\cite{du2020understanding,Zannettou_Finkelstein_Bradlyn_Blackburn_2020,zannettou2018origins} and have been used by state-sponsored actors to wage information warfare~\cite{abidin2020MemeFactoryCultures,zannettou2019characterizing}.
We find evidence of fauxtography images being turned into memes and being manipulated in ways not related to their original verisimilitude.

Finally, by focusing on three selected case studies of fauxtography which spawned new variants, we will discuss implications for dealing with fauxtography in the wild, considering the current environment of social media moderation.

\section{Fauxtography}

The term fauxtography was first coined by~\cite{cooper2007concise} in the context of the 2006 Lebanon war, as combination of the word {\em faux} (French for false) and {\em photography}.
Cooper defines fauxtography as ``visual images, especially news photographs, which convey a questionable (or outright false) sense of the events they seem to depict.''
Fauxtography usually involves manipulated images aiming to influence the emotions of viewers.
Therefore, it involves deception, often realized by directly manipulating the images, captions, or overall the narrative associated with the image.

\begin{figure} \centering
\subfigure[Manipulated]{
\includegraphics[width=0.45\columnwidth]{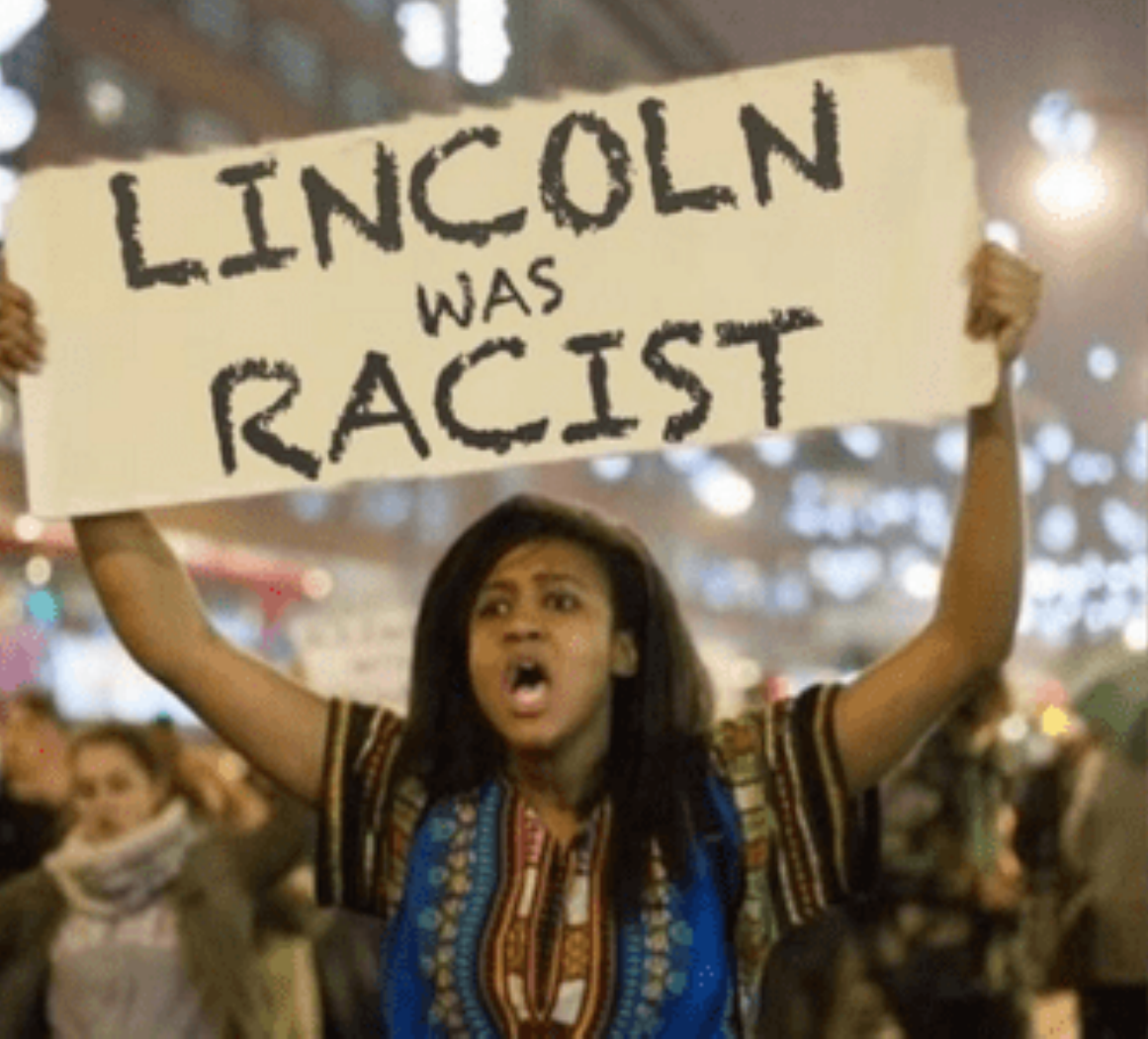}
\label{fig:a}
}
\subfigure[Original]{
\includegraphics[width=0.45\columnwidth]{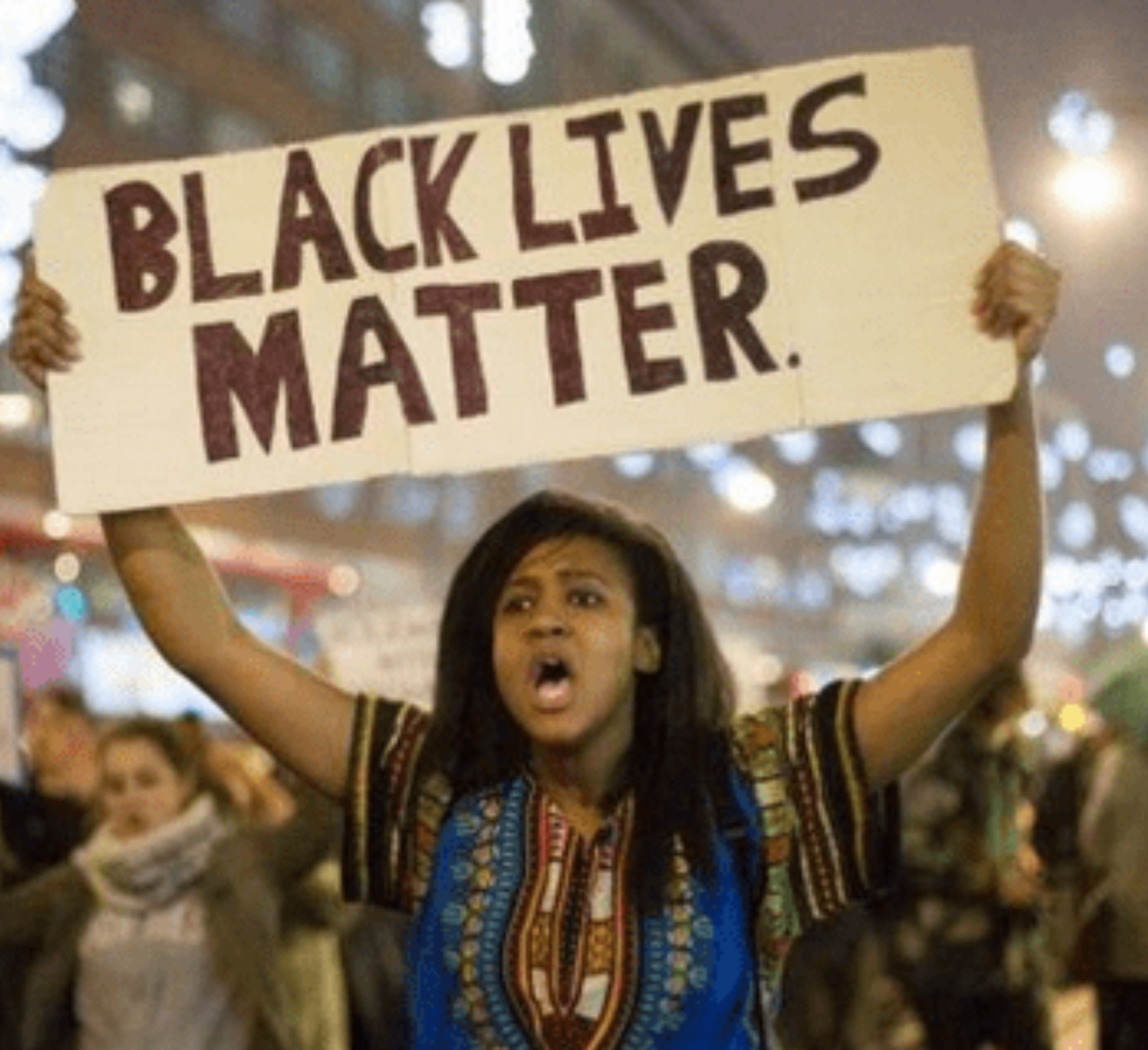}
\label{fig:b}
}
\vspace{-0.35cm}
  \caption{This picture originally depicted a UK protester holding the ``Black Lives Matters'' sign. It was manipulated so that the sign says ``Lincoln was Racist'' and the person has been mischaracterized as being a Missouri State University student. See \url{https://www.snopes.com/fact-check/abe-lincoln-racist-protest-sign/}}%
\label{Racist is actually BLM}
\reduce
\end{figure}

\begin{figure}[t]
  \centering
  \includegraphics[width=0.8\columnwidth]{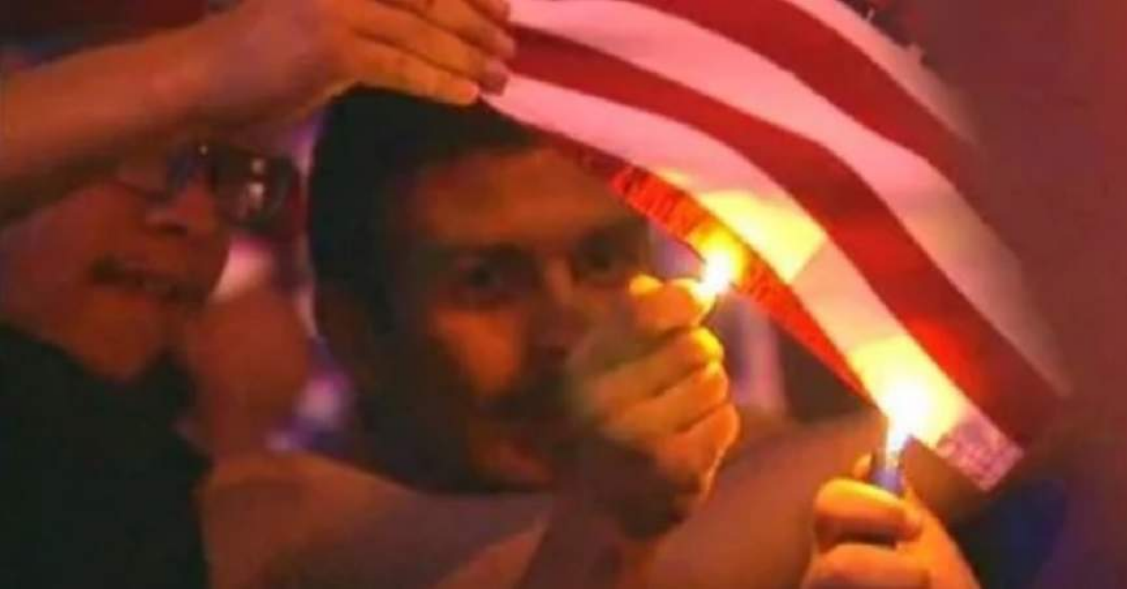}
  \vspace{-0.15cm}
  \caption{Miscaptioned image used to falsely claim that people in the migrant caravan burnt an American flag. See \url{https://www.snopes.com/fact-check/caravan-burning-flag/}}\label{caravan_flag}
  \reduce
\end{figure}

To better explain what fauxtography is, we provide two examples.
Figure~\ref{Racist is actually BLM} shows a picture of a protester in the UK holding a sign reading ``Black Lives Matter,'' which was manipulated to instead read ``Lincoln Was Racist.''
Online sources also erroneously claimed that the person holding the sign was a Missouri State University student at a US protest.
Figure~\ref{caravan_flag} shows an image that was not manipulated, but that has often been used out of context and miscaptioned to imply that migrants on a caravan to the US in 2018 had burned the American flag.
In reality, this photo was taken at an anti-Trump protest in the US and the flag is actually a Trump banner, not a US flag.
These examples demonstrate two important characteristics of fauxtography,  distinguishing them from ``simple'' fake images: 1) they are related to news or public affairs, and 2) users who see them can be fooled relatively easily if the images are not fact-checked. %

\section{Methodology \& Dataset}
\begin{figure}[t]
      \centering
      \includegraphics[width=0.7\columnwidth]{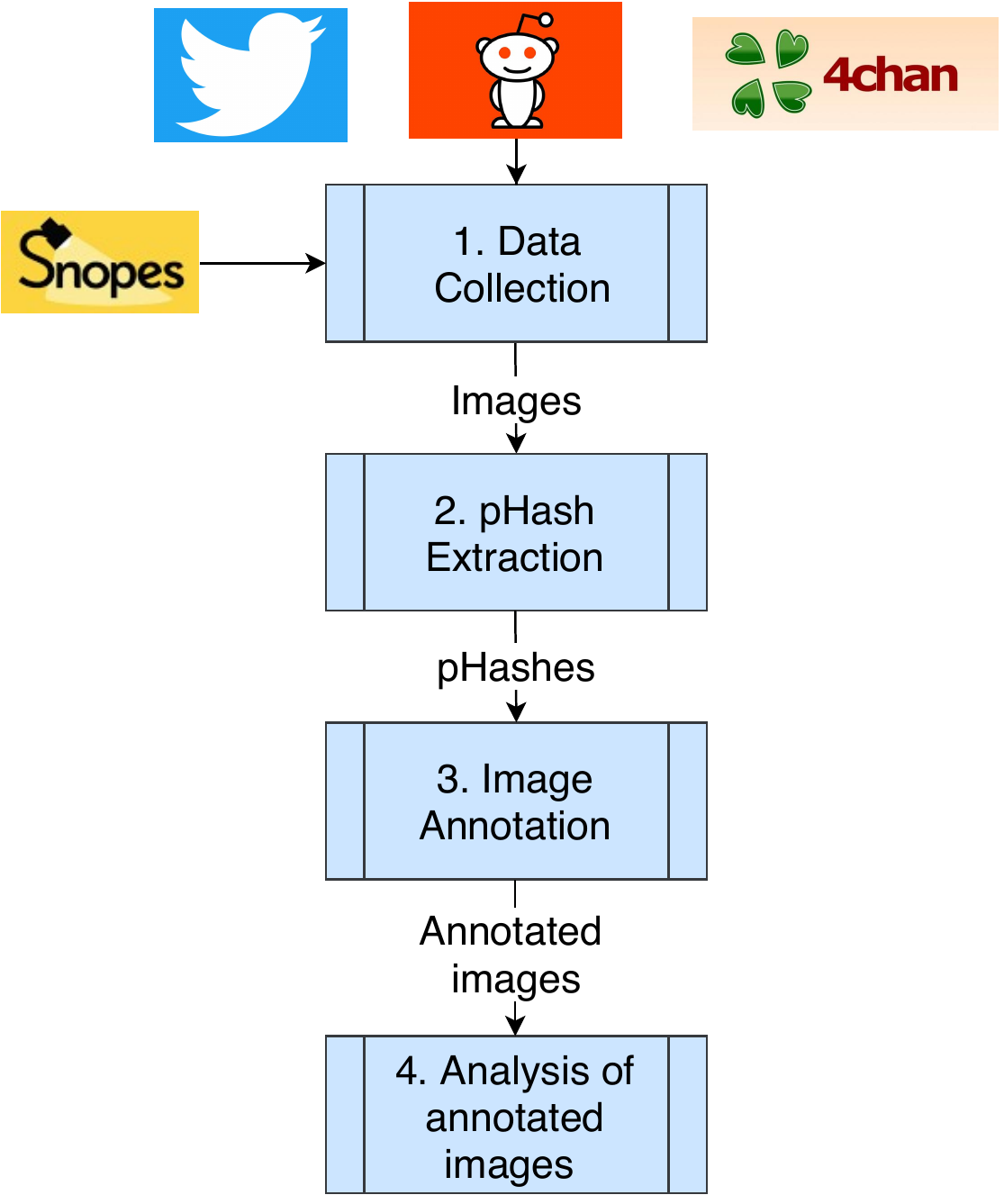}
	\vspace{-0.15cm}
     \caption{Overview of our computational analysis pipeline.}\label{pipeline_image_news} %
\end{figure}

In this section, we present our computational pipeline, as well as the dataset used in our study.
As depicted in Figure~\ref{pipeline_image_news}, the pipeline consists of four components:
1) data collection,
2) pHash extraction,
3) image annotation,
4) image analysis.

\subsection{Data Collection}\label{data_collection}

Our study relies on two types of data sources:
1)~{\em images} from Web communities and news articles posted on them; and
2)~{\em annotation sources} to identify which images are fauxtography.
For the former, we use Twitter, Reddit, and 4chan, and in particular images shared between July 1, 2016 and October 31, 2018; basic statistics are reported in Table~\ref{tbl:datasets}.
As an annotation source, we use \url{Snopes.com}, and specifically images posted on its fauxtography section.
This allows us to identify all images that Snopes classified as fauxtography between the early 2000s and October 2019.
Note that our analysis pipeline supports any Web community and any annotation source; however, in the following, we provide details of the sources used in this paper.

\begin{table}[t!]
\centering
\small
\setlength{\tabcolsep}{4pt}
\begin{tabular}{@{}lrrr@{}}
\toprule
\textbf{Platform}  & \textbf{\#Posts} & \textbf{\#Image} & \textbf{\#Images} \\
& & {\bf URLs} & {\bf Obtained}\\
\midrule
       Twitter &                  {2,213,019,239} &                    {701,806,921} &                     {435,244,799}                      \\   %
        Reddit &                 {295,460,914} &                   {78,682,398} &                    {61,703,316}       \\ %
         4chan &                   {99,614,382} &                      {27,044,132}              &          {23,379,630}    \\
           News Articles &                 {32,200,604} &                   {28,654,146} &                      {27,360,218} \\ %
           Snopes   &                 {2,286} &                  {16,206} &                      {7,835} \\
\bottomrule
\end{tabular}%
\vspace{-0.15cm}
\caption{Overview of our datasets. }
\label{tbl:datasets}
\reduce\reduce
\end{table}

\descr{Images shared on Web communities.} First, we collect images posted publicly on Twitter, 4chan, and Reddit.
For Twitter, we collect data using the 1\% Streaming API, with tweets stored as they were posted, in real time.
In total, we parse 2.2B tweets, 702M of which contain at least one image.
Note that the Twitter API does not return images directly, but rather a URL pointing to the image.
We download the images in March 2020 and are able to retrieve 435M of them.
The remaining images are unavailable, either because the image URL had changed, the tweet was deleted, or because the account that posted it was suspended.

For Reddit, we use the Pushshift dataset~\cite{baumgartner2020pushshift}.
We obtain 295M posts, 79M of which contain images.
Of these, we successfully retrieve 62M images, with the rest having been deleted.
For 4chan, we use the dataset from~\cite{papasavva2020raiders} and obtain 100M posts from 4chan's Politically Incorrect board (/pol/).
The dataset does not include the images posted on /pol/ (only an md5 checksum), hence we use \url{4plebs.org}, an archival service, to collect the images.
Overall, we collect 27M image URLs from 4plebs, from which we are able to download 23M images.

\descr{Images from news articles posted on Web communities.}
On most social networks, when a user shares a news article, the platform often automatically generates a preview for it.
Typically, this includes the main image of the article (i.e., the one appearing on the top).
The preview is important with respect to users' image sharing behavior, thus, we complement our image data collection with images included on news articles shared on Twitter, Reddit, and 4chan.
To do so, we use a systematic approach to create a list of news outlets; we start from the top 30K Majestic~\cite{majestic-million} websites released as of February 2019, and use the  VirusTotal API~\cite{virustotal}, a domain categorization service, to get domains categorized as ``news'' and ``news and media.''
Note that the news outlet labeling given by VirusTotal is not always accurate, e.g., domains like \url{adbusters.org} are incorrectly classified as news outlets.
To solve this problem, we use the NewsGuard API~\cite{newsguardtech} to refine the which domains are actually news outlets, and only select those listed in NewsGuard as of February 2019.
In total, we identify 1,037 news outlets.

We then collect posts containing URLs to the 1,037 news outlets posted on Twitter, Reddit, and 4chan, gathering a total of 32M news articles, with approximately 29M including image URLs.
Note that we only consider the top image from each article, which is the image that appears on top of the article. %
To collect the images, we use the Newspaper3k Library~\cite{newspaper3k} to parse the HTML of the 32M news articles, and then extract the URL of the top image identified by Newspaper3k.
We are able to download 27M images from the 29M image URLs in the news articles.

\begin{table*}[t!]
\centering
\small
\resizebox{\textwidth}{!}{
\setlength{\tabcolsep}{5pt}
\begin{tabular}{|l||  c|    c||    c|c|c||c|c|c|c|c|}
\hline
{\bf Original Labels} & True & Mostly True  &Mostly False&False&Miscaptioned&Legend&Outdated&Satire&Unproven&Mixture\\
\hline

{\bf Our Labels} & \multicolumn{2}{|c||}{\cellcolor{green}Merged True}&\multicolumn{3}{c||}{\cellcolor{red}Merged False}&\multicolumn{5}{c|}{\em Not considered} \\
\hline
\end{tabular}
}
\reduce
\caption{Overview of the fauxtography labels assigned by Snopes and of the grouping that we use for the analysis in this paper.}
\label{tbl:ratings}
\reduce
\end{table*}

\descr{Snopes.} As mentioned, we annotate images using Snopes, a website dedicated to fact-checking news, which has a special section dedicated to fauxtography.\footnote{\url{https://www.snopes.com/fact-check/category/photos/}}
Each entry in this section consists of a topic and a claim associated to an image, which is rated by Snopes using ten possible labels, listed in Table~\ref{tbl:ratings}.
For our analysis, we merge these labels into two groups: Merged True (True and Mostly True) and Merged False (Mostly False, False, and Miscaptioned), as illustrated by Table~\ref{tbl:ratings}. %
The former category includes cases where although part of the claim might be inaccurate, the usage of the image is still correct, whereas, the latter indicates that the usage of an image for a given claim is problematic.

We collect data from the Snopes fauxtography category from the very beginning of the site (early 2000s) to October 2019, obtaining 2,286 articles.
These include 16K URLs to images, out of which we successfully download 7.8K (the rest of the URLs are no longer available).

\subsection{pHash Extraction}\label{sec:phash}
Having collected all images from our data sources, the next step in our pipeline is to convert the images to a format that we can easily work with.
To do so, we apply the Perceptual Hashing (pHash) algorithm~\cite{monga2006perceptual} using the ImageHash library~\cite{imagehash2020}, which generates a hash for each image in such a way that visually similar images have minor differences in their hashes.
The algorithm is robust to image transformations (e.g., slight rotation, skew). %

\subsection{Image Annotation}
\label{subsection:image_annotation}
Next, we annotate and identify the images that relate to fauxtography.
To do this, we perform pairwise comparisons between the pHashes of images obtained from the various Web communities (including news articles) and images obtained from image annotation sites, such as Snopes.
We calculate the Hamming distance between a pair of pHashes (i.e., an image from Snopes and an image shared on Twitter) and we assume that an image is related to fauxtography if the Hamming distance is less than or equal to a pre-defined threshold, which we set below.
Previous work~\cite{zannettou2018origins} shows that pHash is ineffective when dealing with images that are dominated by a single background color (e.g., screenshots on a white background),
thus, %
we remove from our dataset images from annotation sites dominated by a single background color (i.e., screenshots, images of sky, etc.).
Overall, this leaves is with 5,789 Snopes images for subsequent analysis.

\begin{figure}[t]
  \centering
  \includegraphics[width=0.85\columnwidth]{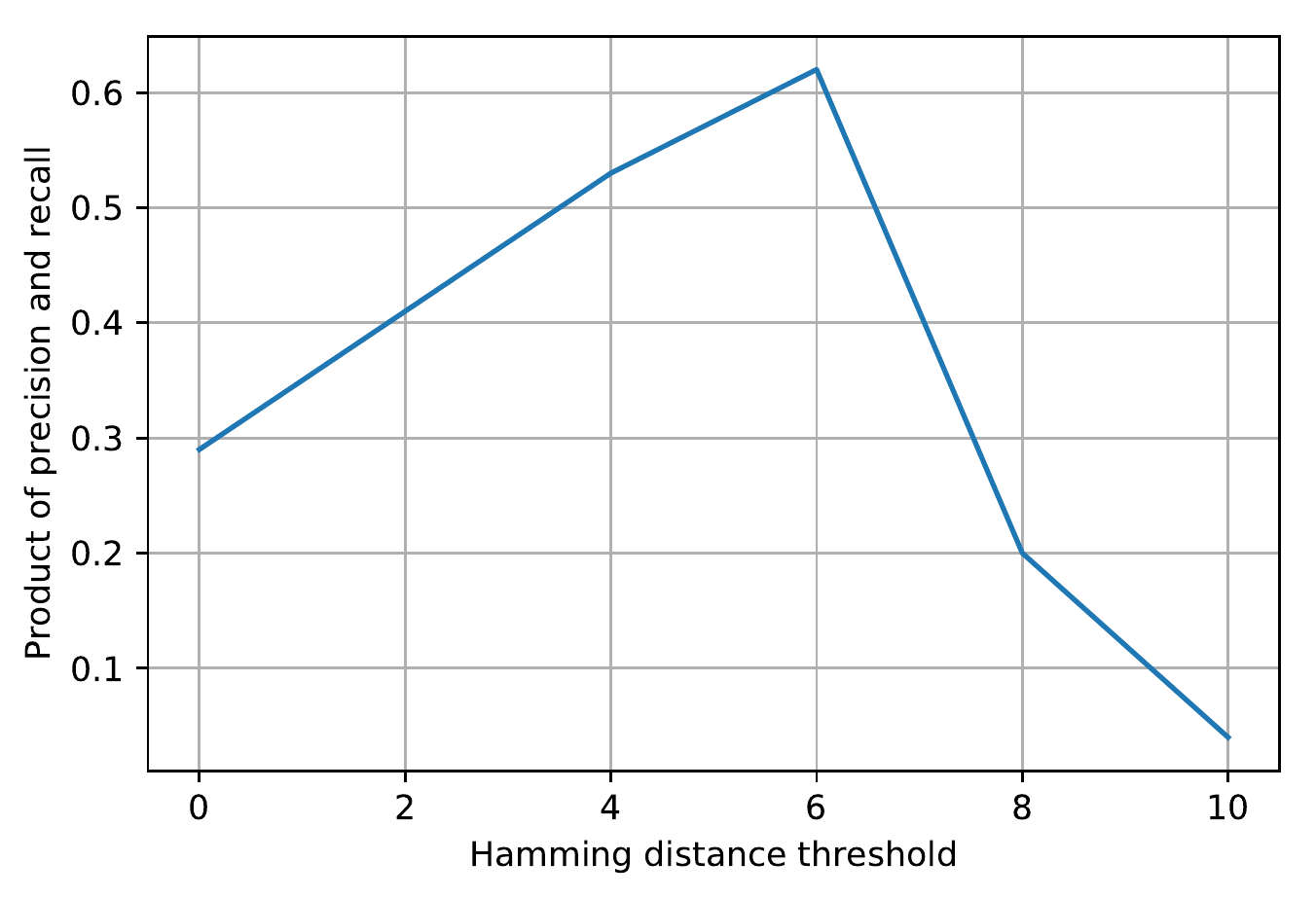}
  \reduce\reduce
  \caption{Product of precision and recall at different pHash Hamming distance thresholds in the image annotation process.}\label{threshold_selection_process} %
  \reduce\reduce
\end{figure}

\descr{Setting the pHash threshold.} %
We consider two images to be visually similar if the Hamming distance is below a certain threshold.
We vary the threshold from 0 to 10 and perform manual inspection of the matched images between the top images of news articles shared on all three Web communities and the corresponding Snopes images.\footnote{Empirically, we find that any pair of images with Hamming distance above 10 consists of extremely dissimilar images.}\bb{Only news article images, or all images?}\yw{only news article images}
We consider a match to be correct if a human annotator considers them visually similar. %
For each value of the Hamming Distance, we calculate the product of precision and recall for all pairs. %
In total, we manually check 76,067 pairs of matched images.

The result of the pHash threshold selection process is shown in Figure~\ref{threshold_selection_process}: the maximum product of precision and recall is obtained at Hamming distance 6 (0.89 precision and 0.69 recall), hence, we use 6 as the threshold to determine if two images are similar.
At this threshold, we find that 2,129 %
fauxtography images from Snopes have at least one match in posts on one of the social networks or in our dataset.
In total, we find 45,567 tweets, 10,916  submissions and comments from Reddit, 2,987 posts from 4chan, and 1,633 news articles that include fauxtography.

\section{RQ1: Impact on Engagement}\label{sec:engagement}

To understand if including fauxtography in social media posts increases engagement, we first look at whether posts on Twitter containing fauxtography produce more retweets and likes than other posts.
We next look at submissions on Reddit, where we use the scores %
 that a submission receives and the length of threads as engagement metrics.
Finally, we look at posts on Twitter and Reddit that do not include fauxtography directly, but that rather include links to news articles containing fauxtography.
Note that we do not analyze the 4chan data here because the small number of data points makes statistical analysis unsuitable (301 threads fauxtography images in total for images that are shared directly, and 38 images for news articles containing fauxtography).\bb{I thought we were going with no good baselines?}

\begin{figure}[t]
\centering
\includegraphics[width=0.7\columnwidth]{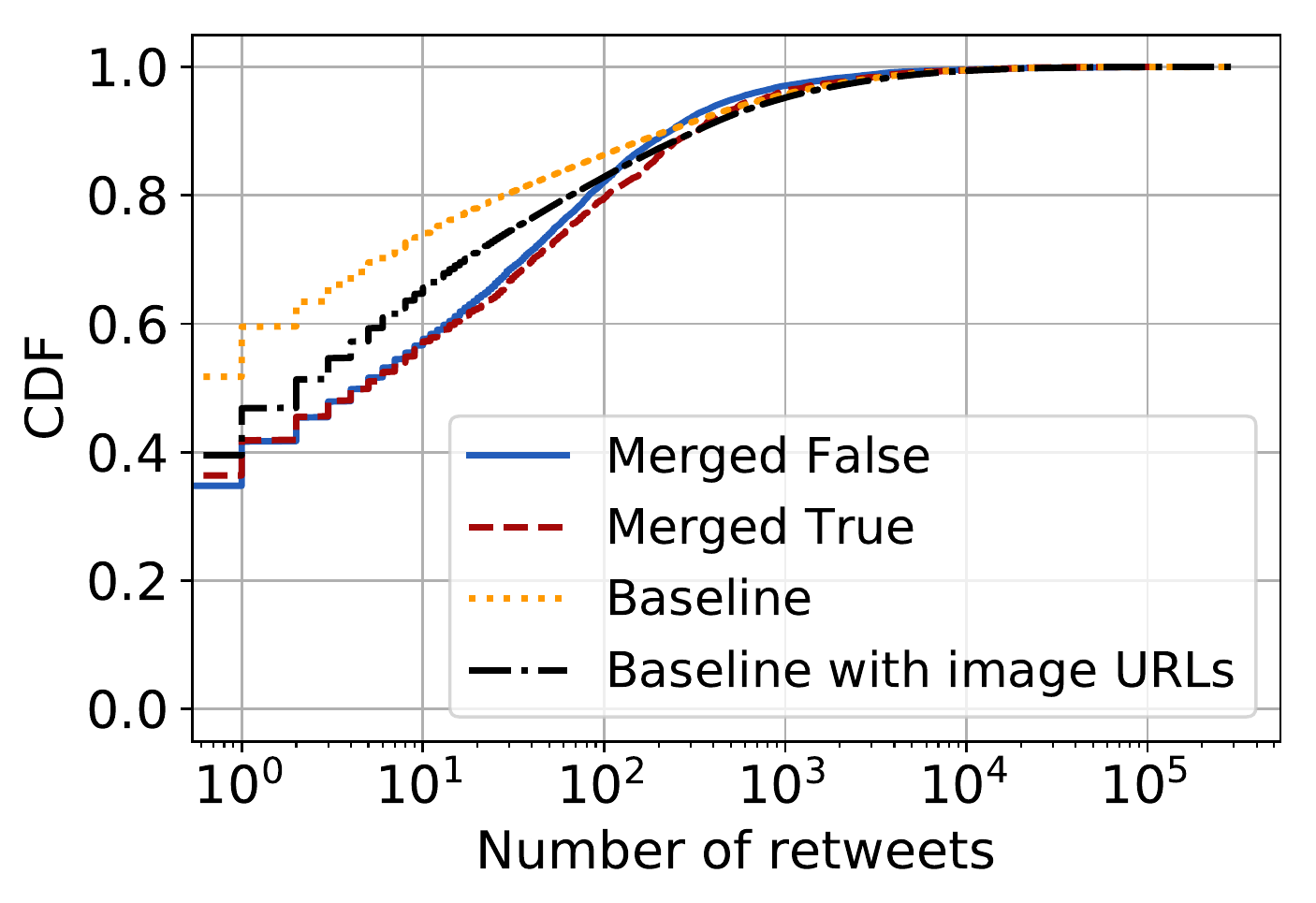}
\caption{CDF of number of retweets on tweets sharing directly images.}
\label{Twitter_retweets_direct_images}
\end{figure}

\begin{figure}[t]
\centering
\includegraphics[width=0.7\columnwidth]{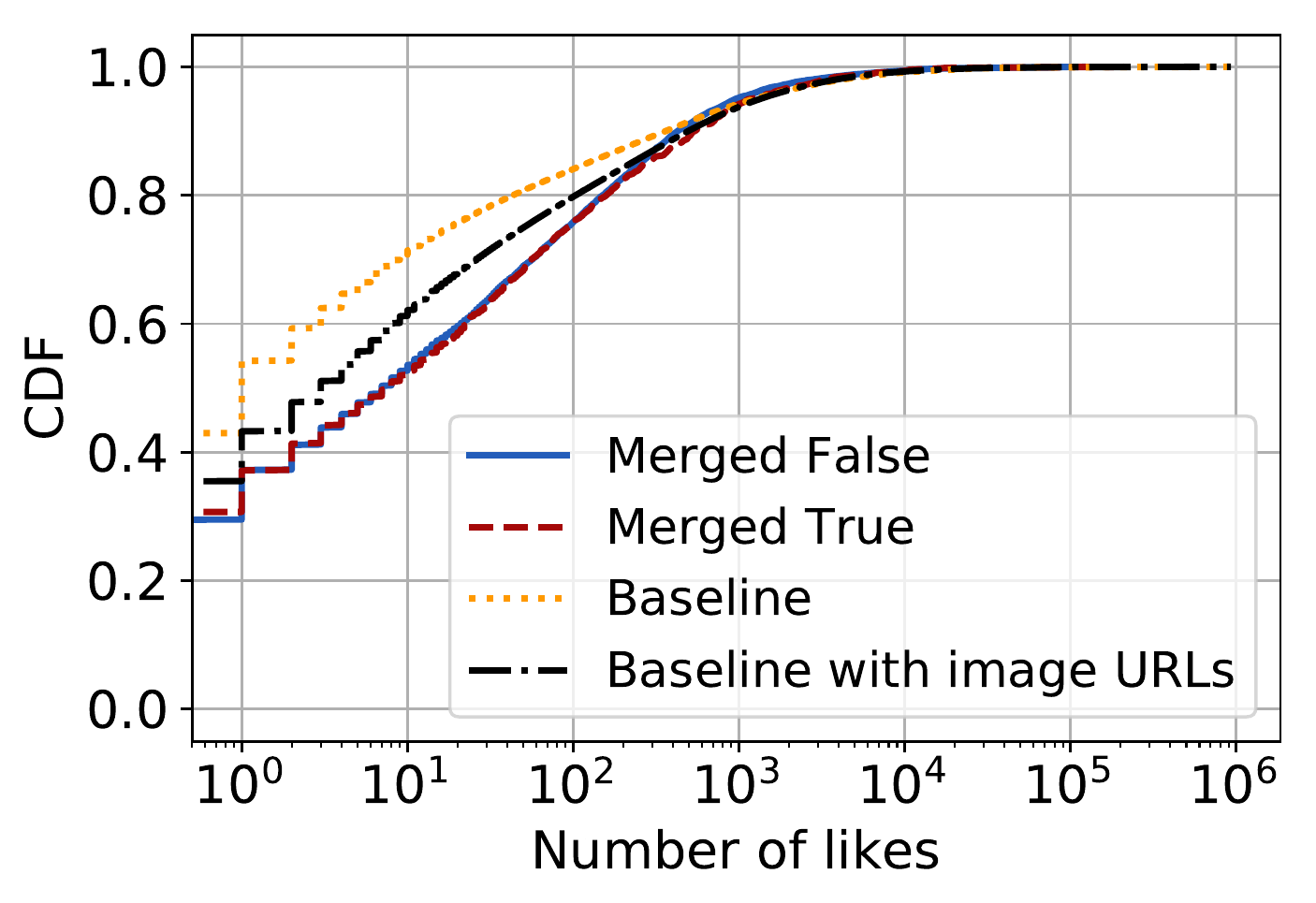}
\reduce\reduce
\caption{CDF of number of likes on tweets sharing directly images.}
\label{Twitter_likes_direct_images}
\end{figure}

\begin{figure}[t]
\centering
\includegraphics[width=0.7\columnwidth]{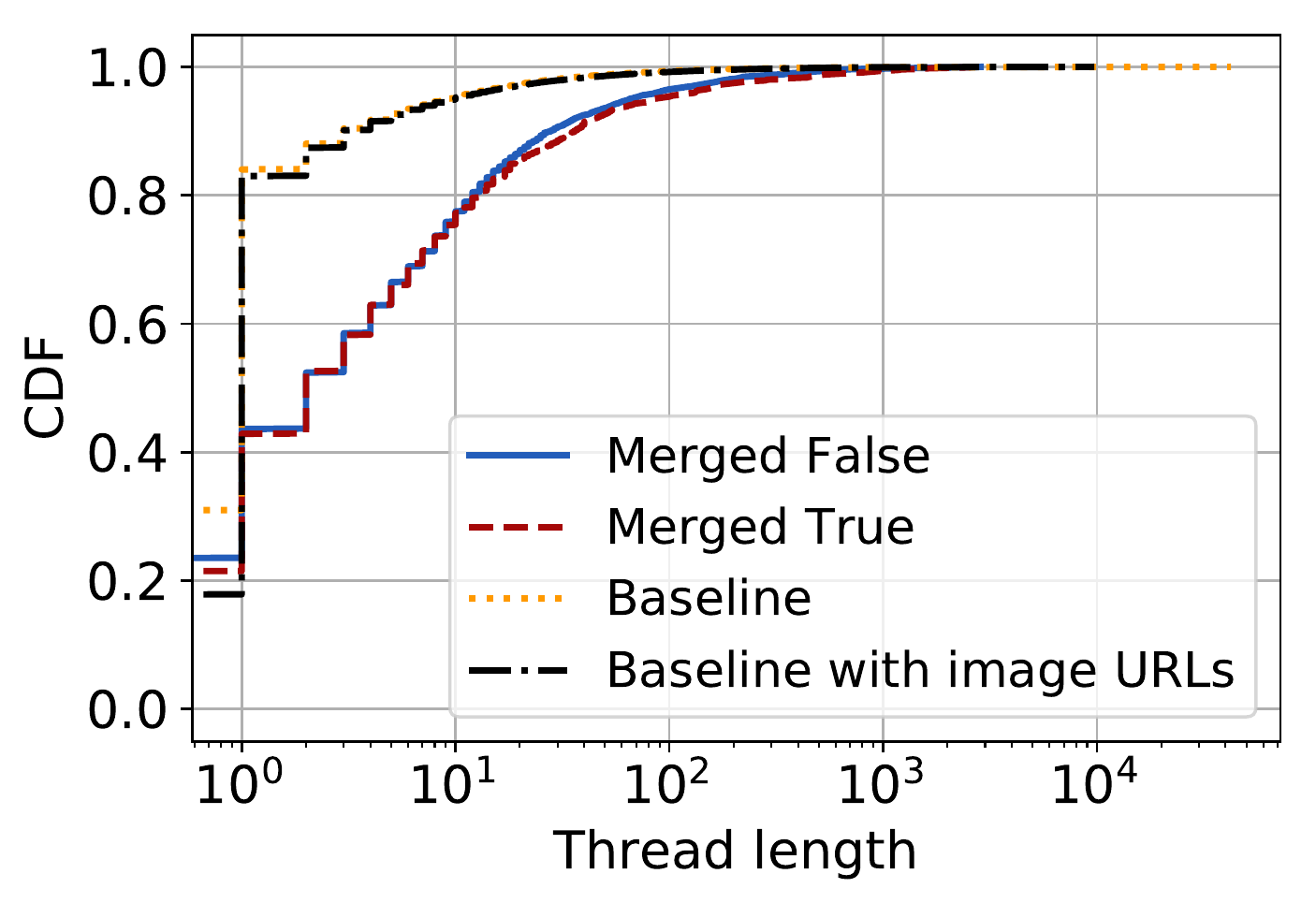}
  \caption{CDF of Reddit submission thread length on submissions sharing directly images.}
  \label{Reddit_threadlength_direct_images}
\end{figure}

\begin{figure}[t]
\centering
\includegraphics[width=0.7\columnwidth]{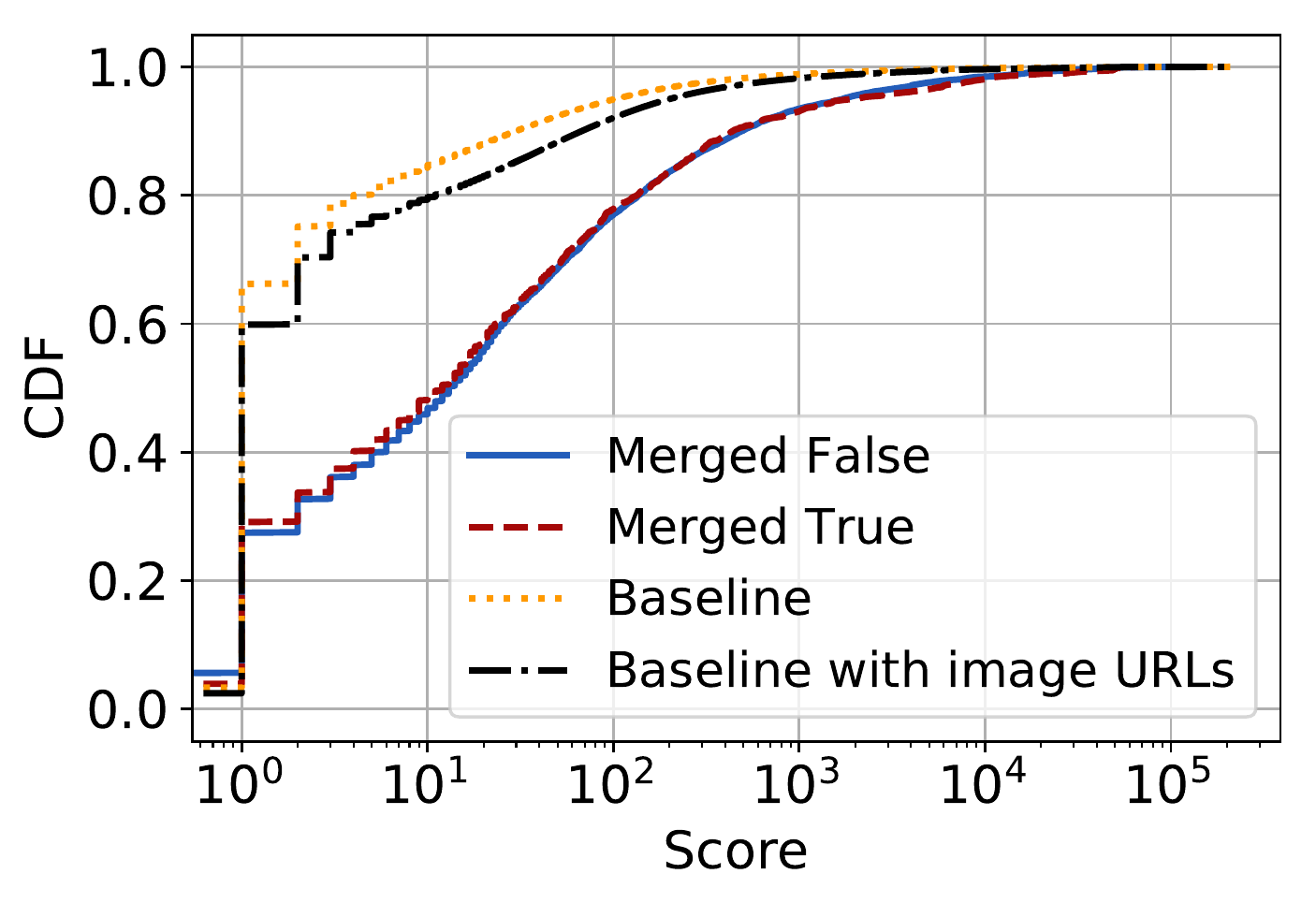}
  \caption{CDF of Reddit submission score on submissions sharing directly images.}
  \label{Redditlikes_direct_images}
\end{figure}

\subsection{Twitter}
As we use the Twitter streaming API for data collection,  our data contains real time activity, i.e., tweets are gathered as soon as they are posted.
This makes the dataset less than ideal to assess the engagement received by tweets, because the number of retweets and likes reported by the API represents short-term behavior.
To gain a better view of the long-term engagement, we leverage a process called \emph{hydration\footnote{\jbnote{WE CAN SAVE SPACE IF NEEDED BY MOVING THESE TO REFERENCES PROBABLY}\url{https://developer.twitter.com/en/docs/twitter-api/v1/tweets/post-and-engage/api-reference/get-statuses-lookup}}}: given a tweet ID, we retrieve the latest version of the current number of retweets and likes for that tweet.
We hydrate the tweets in our dataset between June and July 2020.

Tweets can be classified as original tweets, retweets, and quote tweets. %
After hydration, we find that we cannot retrieve the actual retweet and likes count of regular retweets.\footnote{The  ``retweet\_count'' field in the metadata of the retweet, representing how many times a tweet is retweeted, is always equal to the ``retweet\_count'' field in the metadata of the corresponding original tweet.
In addition, the field ``favorite\_count,'' i.e., how many times a tweet is liked,  is always 0 in the metadata of a retweet even if users press ``like'' on the retweet instead than on the original tweet, and the ``favorite\_count'' of the original tweet is increased instead.} %
Therefore, to assess engagement for retweets, we retrieve the latest version of the original tweet that generated the retweet.
As discussed earlier, Snopes provides detailed labels to characterize fauxtography.
For our experiments, we combine similar ratings together and form a binary system with two classes, Merged True and Merged False (see Table~\ref{tbl:ratings}).

To understand whether posts containing fauxtography produce more engagement on Twitter, we extract two baselines: a set of random tweets and a set of tweets containing images that are not labeled as fauxtography.
We then compare the engagement distribution of these tweets to posts containing fauxtography.
We identify 9,858 Twitter users that shared tweets containing Merged True and Merged False fauxtography.
\yw{This sentence can be modified as ``By hydration the tweets containing fauxtography, we are able to obtain 28,982 existing tweets. From which, we identify 9,858 users and 9,771 and 2,183 non-retweeted tweets containing merged false and merged true fauxtography, respectively. ''}
We collect 9,771 tweets\sz{in Section 3, we report 50K tweets. which one is the correct?}
\yw{the numbers in section 3 containing merged true, merged false, and other ratings. In addition, the number here is obtained after redownloading tweets. Many tweets have already been unavailable.} containing fauxtography rated as Merged False, 2,183 tweets containing fauxtography rated as Merged True.
Then, we construct two baselines deriving from all the tweets shared from these 9,858 Twitter users:
1) 1,720,197 tweets that do not include images;
and 2) 782,391 tweets that include non-fauxtography images.

Figures~\ref{Twitter_retweets_direct_images} and~\ref{Twitter_likes_direct_images} show the cumulative distribution functions (CDFs) of the retweets and likes received by the four types of tweets, respectively.
We observe that tweets containing an image from our fauxtography dataset (whether true or false) are more likely to produce more retweets and likes than our baseline tweets:
$42\%$ tweets containing Merged False fauxtography and  $43\%$ tweets containing Merged True fauxtography have been retweeted more than 10 time, while only $26\%$  tweets from the generic baseline of random tweets have been retweeted more than 10 times.
\yw{we can change the sentence to `` the median of retweets received by tweets containing Merged False fauxtography and tweets containing Merged True fauxtography are both  5 while the median for the times that generic baselines have been retweeted is 0.''}\bb{At this point, lets go with whatever phrasing is shortest.}
Similarly, $46\%$ tweets containing Merged False fauxtography and $47\%$ tweets containing Merged True fauxtography have been liked more than 10 times, while only $29\%$ tweets from the generic baseline have been liked more than 10 times.
\yw{we can change the sentence to `` the median of likes received by tweets containing Merged False fauxtography and tweets containing Merged True fauxtography are 7 and 8, respectively, while the median for the times that generic baselines have been liked is only 1.''}\
\bb{Why the focus on 10 times? Why not report, e.g. the median number of likes/retweets for each curve?}
\yw{@Barry, I changed above.}

To assess differences between these distributions, we run two sample Kolmogorov-Smirnov tests (K-S test)~\cite{lindgren1993statistical}.
We first compare to the baseline set of random tweets. We find that the differences between the following distributions are statistically significant at the $p<0.05$ level: Merged False tweets compared to the baseline ($D{=}0.182$), and Merged True tweets compared to the baseline ($D{=}0.185$) when examining retweets.
As for likes, we have statistically significant differences between the distribution of Merged False tweets compared to the baseline ($D{=}0.187$), and for Merged True tweets compared to the baseline ($D{=}0.192$). In all cases, $p\ll 0.001$
We thus reject the null hypothesis that tweets with fauxtography images receive the same level of engagement as random tweets.

\jbnote{this and the previous paragraph seem to have results kinda mixed up... need to check previous paragraph and make sure that it's not talking about the same stuff in this paragraph}\bb{I think these two paragrpahs should be restructured to deal with this. The problem is the previous paragraph introduced all the baselines rather confusingly, and then ignored half of them. I think its worthwhile to only introduce the baselines as they are needed.}

There is reason to believe that tweets containing images get more engagement overall~\cite{li2020picture}.
To lend further evidence to the observation that our images in our fauxtography dataset are likely to receive more engagement than random images, we next compare the fauxtography distributions to the baseline of tweets with images in Figures~\ref{Twitter_retweets_direct_images} and~\ref{Twitter_likes_direct_images}.
Again, we observe that tweets with images from our fauxtography dataset are more likely to be retweeted and liked than those with other images: only $34\%$ of tweets containing non-fauxtography image have been retweeted more than 10 times, and only $38\%$ have been liked more than 10 times.
\yw{we can change the sentence to ``From the figures, we see that again, tweets with images from our fauxtography dataset are more likely to be retweeted and liked than those with other images:  the median of retweets received by tweets containing non-fauxtography image is 2, and the median for the likes is 3.''}
Using 2-sample K-S tests, we reject the null hypothesis that tweets containing non-fauxtography images and those with fauxtography have the same probability of receiving engagement ($p \ll 0.001$ in all cases).
For retweets, we have $D{=}0.0811$ for Merged False tweets compared to non-fauxtography image baseline, and $D{=}0.0896$ for Merged True tweets compared to non-fauxtography image baseline.
For likes we have $D{=}0.0854$ for Merged False tweets compared to the non-fauxtography image baseline, and $D{=}0.0968$ for Merged True tweets compared to the non-fauxtography image baseline.\sz{we are using different terms than the ones used in the figures for the baselines. better be consistent?}
Finally, a question remains as to whether or not the verisimilitude
of a fauxtography image affects its engagement.
We compare the distribution of engagement between tweets with Merged True and Merged False images.
In this case, we reject the null hypothesis that there is a difference with respect to retweets ($D{=}0.0380$, $p = 0.011$), however we are \emph{unable} to reject the null hypothesis of differences with respect to likes ($D{=}0.0219$, $p = 0.36$).
One explanation for this result is that images in our fauxtography dataset are usually quite controversial, with a sensationalist tone. We speculate that this tends to drive engagement, regardless of the underlying verisimilitude%
of the image itself.

\subsection{Reddit}

For Reddit, we run analogous experiments using the length of a thread and the score of a submission as engagement metrics.
Reddit calculates the score of a post as the difference between the number of upvotes and downvotes that it receives.\gs{check that this is accurate}\jbnote{added a note that there is some score fuzzing that goes on on reddit. we should probably find a cite}
On Reddit, the initial post in a thread is the ``submission,'' and other posts in that thread are ``comments.''
The length of a thread %
is obtained from the ``num\_comments'' metadata field, and the score (i.e., the number of upvotes minus the number of downvotes) is obtained from the ``score'' field in submission metadata.\bb{Can we say ``the length and score for each thread is contained in the dataset?''}
Note that the ``score'' field is a precise value\footnote{\url{https://www.reddit.com/wiki/faq\#wiki_how_is_a_submission.27s_score_determined.3F}} while upvote and downvote values are fuzzed.

First, we identify 4,883 users that shared submissions containing fauxtography.
These users shared 5,444 submissions containing Merged False fauxtography and 1,522 submissions  containing Merged True fauxtography, respectively\jbnote{are these UNIQUE images? i.e., how many snopes images does this account for?}.
\yw{We use 847 Snopes image URL in total. 645 belongs to merged false and 202 belongs to merged true}
\gs{how can the number of fauxtography submissions be lower than the number of users sharing fauxtography submissions?}\bb{+1}\yw{changed in the text, it should be ok now}
Then, we construct two baselines based on the same set of Reddit users:
1)~7,248,595 submissions that do not include images;
and 2)~3,367,222 submissions that include non-fauxtography images.

Figures~\ref{Reddit_threadlength_direct_images} and~\ref{Redditlikes_direct_images} plot the CDF of thread length and score (respectively) for each of the four sets of submissions just described.
From the plots, we note that $23\%$ of submissions containing Merged False or Merged True fauxtography resulted in threads with more than 10 comments, while this is true for only $4.6\%$ of non-image submissions and $4.8\%$ submissions containing non-fauxtography images.
\yw{change to ``From the plots, we note that the median of thread lengths of submissions containing Merged False fauxtography and  submissions containing Merged True fauxtography are both 2,  while the median of thread lengths of non-image submissions and  submissions containing non-fauxtography images are both 1.''}
Similarly, $53\%$ submissions containing Merged False fauxtography and $53\%$ submissions containing Merged True fauxtography have scores higher than 10, but only $15\%$ of generic non-image submissions and $20\%$ submissions containing non-fauxtography images have a score above 10.\bb{see above note about the choice of 10} %
\yw{change to ``Similarly the median of scores of submissions containing Merged False fauxtography and  submissions containing Merged True fauxtography are 13 and 12, respectively,  while the median of scores of non-image submissions and  submissions containing non-fauxtography images are both 1 as well.''}\bb{pick whichever is shorter, consistent with above}
This suggests that fauxtography images produce more engagement than the baseline, regardless of whether the random post contains an image or not.

The differences in these distributions are again statistically significant as confirmed via  2-sample K-S tests.
For the length of threads, we have $D{=}0.404$ for Merged False submissions compared to the no-image baseline, and $D{=}0.411$ for Merged True submissions compared to the no-image baseline.
For likes, we have $D{=}0.426$ for Merged False submissions compared to the no-image baseline, and $D{=}0.414$ for Merged True submissions compared to the no-image baseline.
When comparing to the non-fauxtography image baseline, we have $D{=}0.394$ for Merged False submissions and $D{=}0.401$ for Merged True submissions when looking at the length of threads.
For likes, we have $D{=}0.381$ for Merged False submissions compared to the non-fauxtography image baseline, and $D{=}0.368$ for Merged True submissions compared to the non-fauxtography image baseline. In all cases, we find $p\ll 0.001$

Similar to Twitter, we are \emph{unable} to reject the null hypothesis that there is no difference in engagement between true and false fauxtography images on Reddit.
In the case of Reddit it is important to note that we are unable to reject the null hypothesis for both types of engagement; $D = 0.0220$, $p = 0.61$ for thread length and $D = 0.0225$, $p = 0.51$ for submission score. Again, this suggests that the engagement generated by fauxtography images is independent of the verisimilitude of the image.
\begin{figure*}
\begin{minipage}[t]{0.24\textwidth}
  \includegraphics[width=\columnwidth]{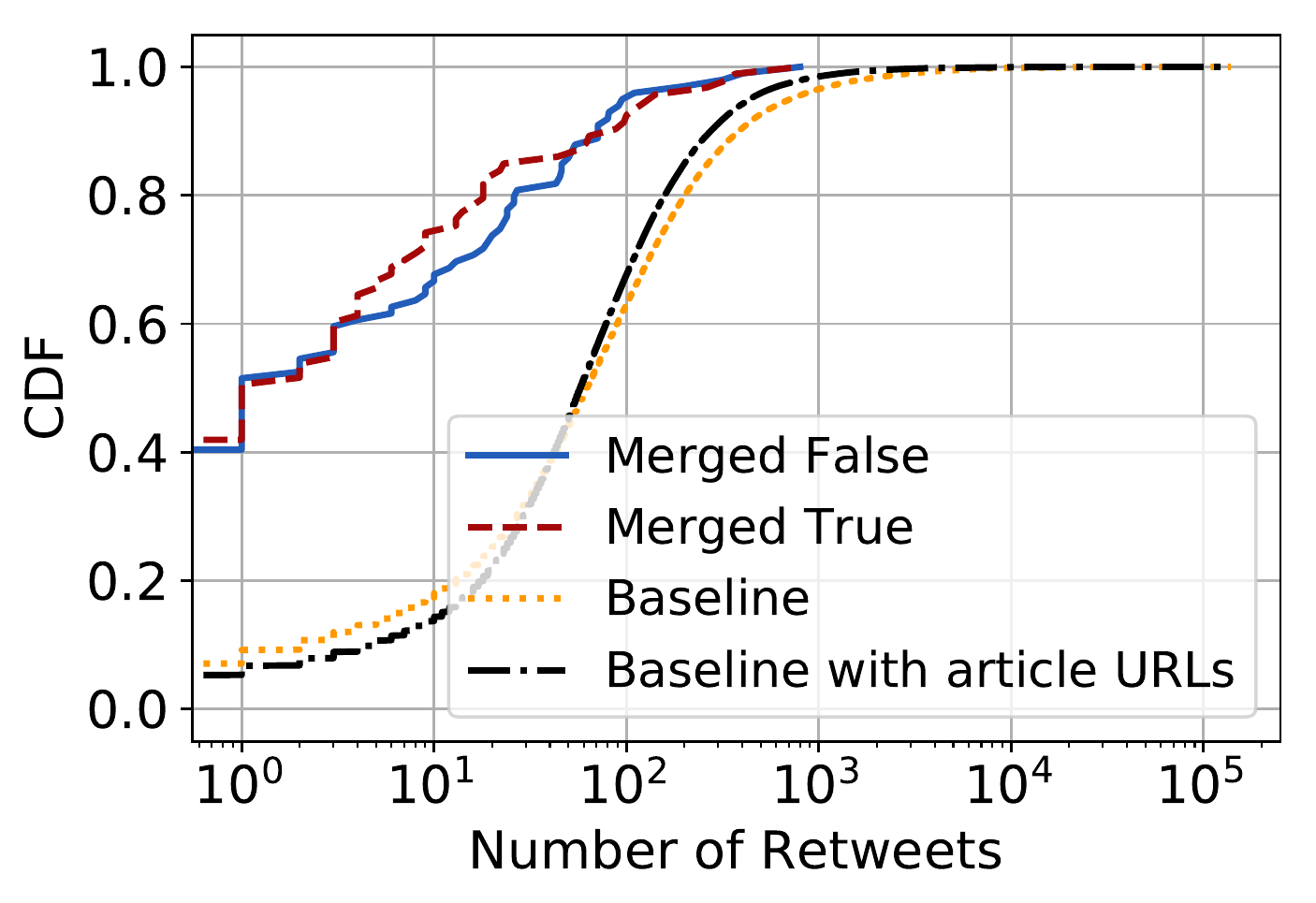}
  \caption{CDf of number of retweets on tweets sharing news articles.}
  \label{fig:twitter_retweets_new_urls}
\end{minipage}
  \hspace{0.02in}
\begin{minipage}[t]{0.24\textwidth}
  \includegraphics[width=\columnwidth]{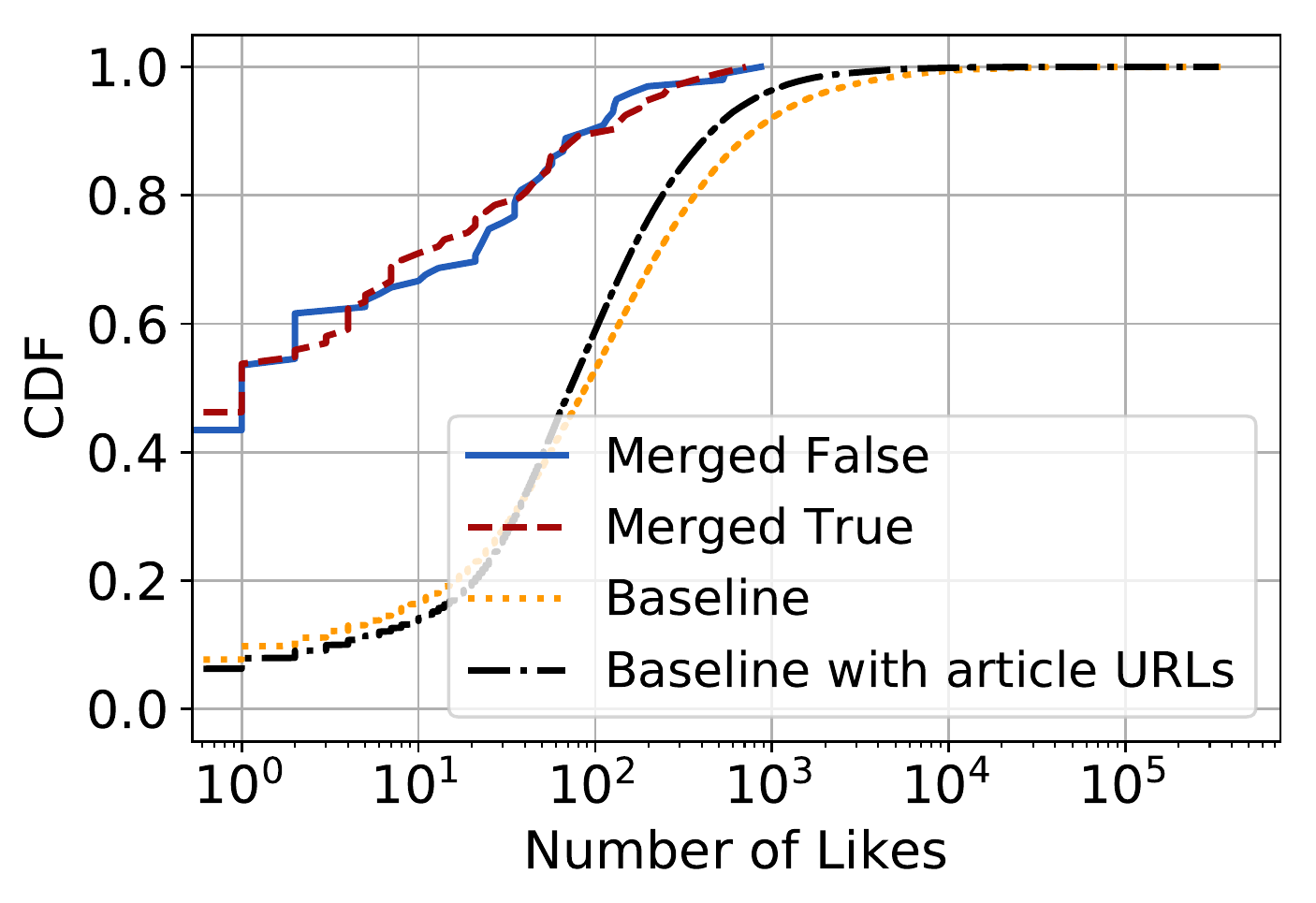}
  \caption{CDf of number of likes on tweets sharing news articles.}
  \label{fig:twitter_likes_news_urls}
\end{minipage}
  \hspace{0.02in}
\begin{minipage}[t]{0.24\textwidth}
  \includegraphics[width=\columnwidth]{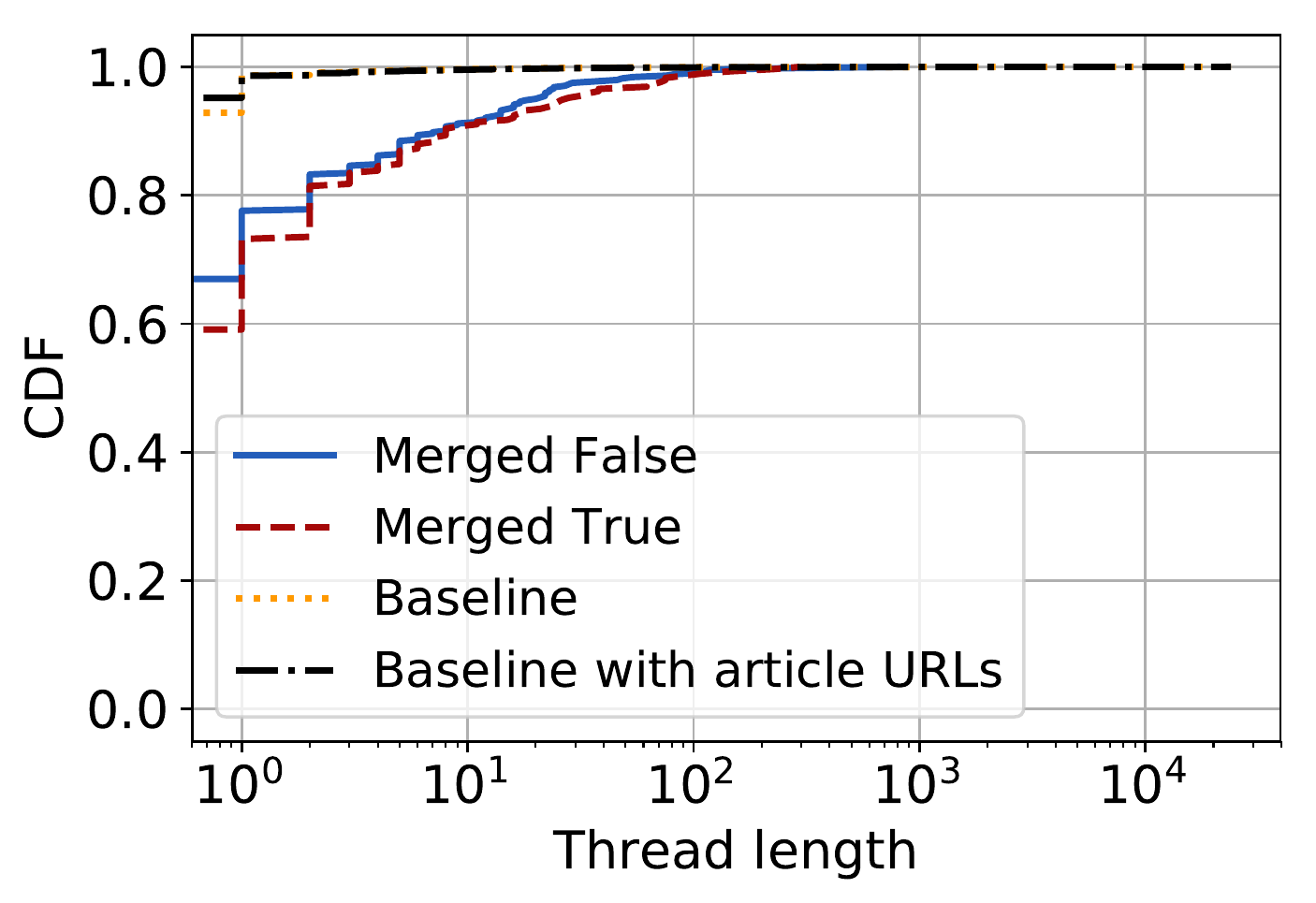}
  \caption{CDF of Reddit submission thread length on submissions sharing news articles.}
  \label{fig:reddit_thread_length_news_urls}
\end{minipage}
\hspace{0.02in}
\begin{minipage}[t]{0.24\textwidth}
  \includegraphics[width=\columnwidth]{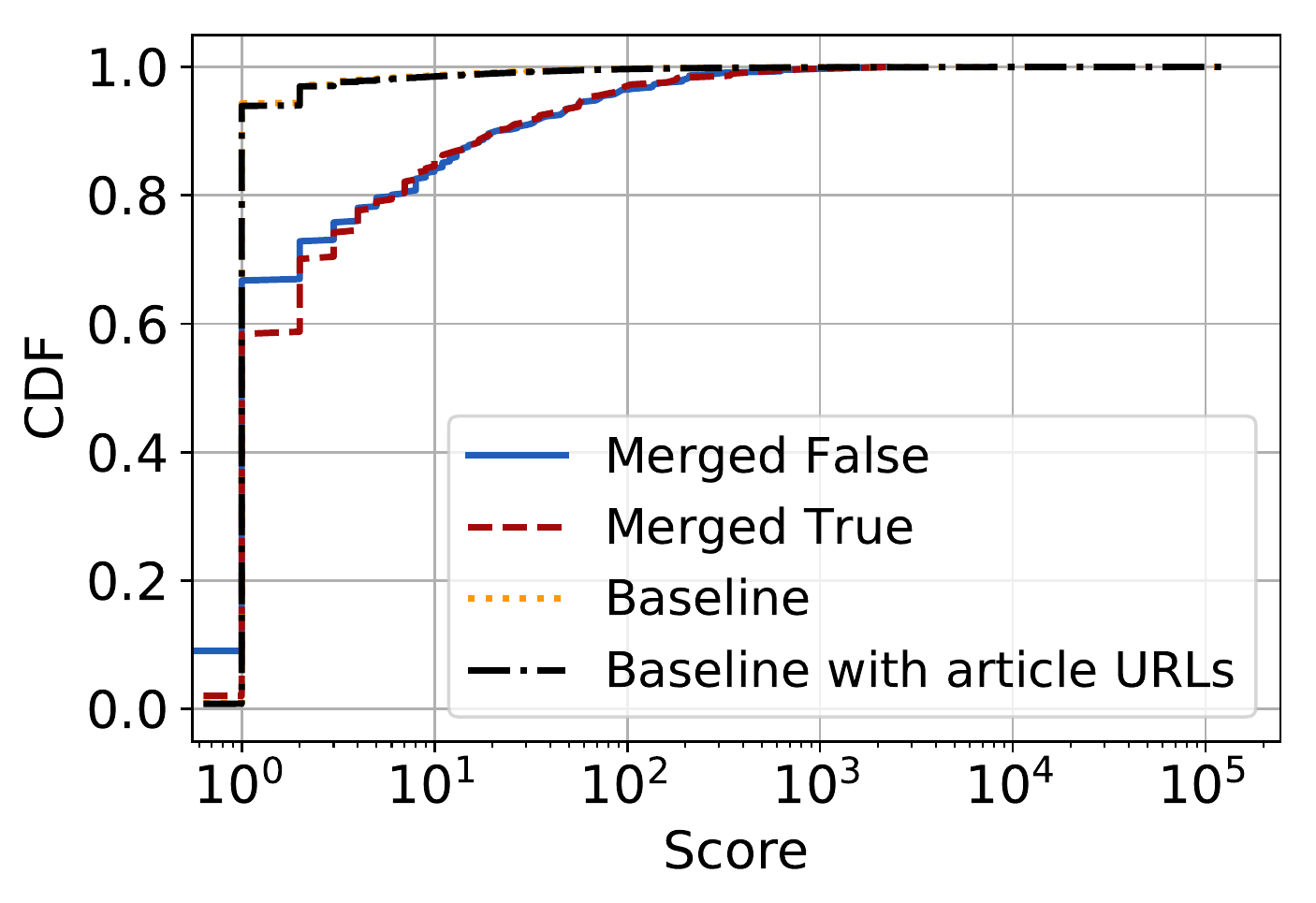}
  \caption{CDF of Reddit submission score on submissions sharing news articles.}
  \label{fig:reddit_scores_news_urls}
  \end{minipage}
\end{figure*}

\subsection{News URLs}

We now look at the engagement generated by posts that %
have links to news articles that include fauxtography rather than directly including fauxtography.
On Twitter, we identify 100 tweets with links to news articles that contain Merged False\gs{make sure that the casing of merged false and true is consistent throughout the paper} \sz{fixed everything on this}\yw{consistent}
fauxtography images and 94 tweets with links to articles that contain Merged True fauxtography images. \yw{after hydration between June and July 2020.}\yw{these tweets are obtained from 176 users.}%
\sz{i think we are trolling in this section. above we said that we excluded 4chan because we only have 300 threads with fauxtography and we cannot get reliable statistical results, however, here we analyze 194 tweets.}\bb{i have to agree here. Unless there are statistics later that show things are still significant. but then we need to show similar stats for 4chan imo}%
On Reddit, we identify 431 submissions with links to articles that contain Merged False fauxtography and 272 submissions with links to articles that contain Merged True fauxtography.
\yw{The Reddit submissions  are from 312 users.}

Once again, we compare the engagement of posts containing links to news articles containing fauxtography to a generic baseline of 492,604 tweets on Twitter and 19,704,911 Reddit submissions, respectively, and to a baseline of 239,079 tweets and 9,554,016 posts containing generic news URLs.
The baselines are constructed by collecting all non-fauxtography posts posted by the users who made at least one fauxtography related submission on Twitter or Reddit. %
Figures~\ref{fig:twitter_retweets_new_urls} and~\ref{fig:twitter_likes_news_urls} show the retweets and likes of tweets containing fauxtography news URLs.
Contrary to what observed previously, these tweets do not receive more engagement than baselines.
More precisely,  on Twitter, $32\%$ tweets containing Merged False fauxtography and $26\%$ tweets containing Merged True fauxtography have been retweeted more than 10 times, while $82\%$ generic tweets and $85\%$ tweets containing non-fauxtography news URLs\bb{Check this} been retweeted more than 10 times.
\yw{change to ' on Twitter, the median times that tweets containing Merged False fauxtography and   tweets containing Merged True fauxtography get retweeted are both 1, while the median times that generic tweets and tweets containing non-fauxtography news urls been retweeted are 62 and 58, respectively.'}
Furthermore, only $33\%$ tweets containing fauxtography rated as Merged False and  $29\%$ tweets containining  fauxtography rated as Merged True have been liked more than 10 times,  while $83\%$ of generic tweets and $86\%$ tweets contain generic  non-fauxtography news URLs\jbnote{generic images or non-fauxtography news urls?}\yw{generic non-fauxtography news urls} have  been liked more than 10 times.
\yw{change to '  the median times that tweets containing Merged False fauxtography and   tweets containing Merged True fauxtography get liked are both 1, while the median times that generic tweets and tweets containing non-fauxtography news URLs been liked are 88 and 72, respectively.'}

On Reddit, on the other hand, we find that posts containing links to fauxtography news articles still receive more engagement.
As Figures~\ref{fig:reddit_thread_length_news_urls} and~\ref{fig:reddit_scores_news_urls} show, $16\%$ of submissions containing Merged False fauxtography and $14\%$ of submissions containing Merged True fauxtography have thread lengths longer than 10, while the same is true only for $1.3\%$ of generic submissions and $1.6\%$ non-fauxtography news URL submissions.
\yw{using median is not good here. because the four medians here are all 1. if we use mean here, then it can be stated as `` the average thread length   of   submissions containing Merged False fauxtography and  submissions containing Merged True fauxtography are 21.0 and 21.3, respectively, while the average thread length  of the generic submissions and   non-fauxtography news URL submissions are 6.5 and 7.3, respectively.''}
\yw{if we want to add scores for Reddit news URLs, then ``the average score   of   submissions containing Merged False fauxtography and  submissions containing Merged True fauxtography are 5.7 and 6.3, respectively, while the average score of the generic submissions and   non-fauxtography news URL submissions are 0.52 and 0.64, respectively.''}\yw{if we use mean here, do we need to change all the median used above to mean to make it consistent.}

\edc{do we really need this level of details of KS tests? this takes a whole column but not sure it's worth it...}
On Twitter, we confirm differences in these distributions via the 2-sample K-S test for fauxtography submissions compared to baseline submissions, where for the number of retweets we have $D{=}0.506$ for Merged False tweets compared to the non-fauxtography baseline, and $D{=}0.584$,for Merged True tweets compared to the generic baseline.
For likes, we have $D{=}0.509$ for Merged False tweets compared to the generic baseline, and $D{=}0.545$ for Merged True tweets compared to the generic baseline.
Looking at the non-fauxtography news article baseline, we have $D{=}0.536$ for Merged False tweets and $D{=}0.614$, for Merged True tweets when looking at retweets.
For likes, we have $D{=}0.534$ for Merged False tweets, and $D{=}0.571$ for Merged True tweets. In all cases, $p\ll0.001$ leading us to reject the null hypothesis that there are no differences between these distributions.

On Reddit, when looking at the length of threads we have $D{=}0.275$ for Merged False submissions compared to the generic baseline, and $D{=}0.358$ for Merged True submissions compared to the generic baseline.
For Scores, we have $D{=}0.260$ for Merged False submissions compared to the generic baseline, and $D{=}0.339$ for Merged True submissions compared to the generic baseline.
When looking at the non-fauxtography news article baseline, we have $D{=}0.271$ for Merged False submissions and $D{=}0.354$ for Merged True submissions for the length of threads.
For Scores, we have $D{=}0.282$ for Merged False submissions compared to the non-fauxtography image baseline, and $D{=}0.362$ for Merged True submissions compared to the non-fauxtography image baseline. In all cases, $p\ll0.001$ leading us to reject the null hypothesis that there are no differences between these distributions.

Note, however, we are unable to reject the null hypothesis that there are differences in engagement between Merged True and Merged False tweets and submissions. %
On Twitter, a KS test gives us $D{=}0.0998$ ($p=0.7$) for retweets and $D{=}0.0562$ ($p\approx 1.0$) for likes. \bb{Changed to approx. Please verify that its .9999 or something and not exactly 1.}
On Reddit, we obtain $D{=}0.0826$  ($p=0.17$) for the length of threads and $D{=}0.0791$ ($p=0.21$) for scores.
While most of the results for this experiment are consistent with what we previously found with regards to directly sharing fauxtography, interestingly, tweets containing links to news articles with fauxtography attract less engagement than other news links.
One possible reason is that when sharing news URLs, many confounding factors can come into play with regards to enticing users into interacting with the tweet, for example clickbait titles and the content of the article. %
For Reddit, the results show that using news articles to share fauxtography can increase engagement, which is consistent with the results found sharing images directly.

\subsection{Takeaways}

Our analysis provides evidence that posts directly containing fauxtography images do indeed generate higher engagement on both Twitter and Reddit.
However, when it comes to sharing links to news articles that make use of fauxtography, we find that they generate significantly \emph{less} engagement on Twitter but significantly \emph{more} engagement on Reddit.\bb{need to reassess in light of suggested changes above.}
Further, except in the case of retweets on Twitter, we are unable to reject the null hypothesis that Merged True and Merged False posts containing fauxtography images or links to news stories using fauxtography receive the same levels of engagement.
Twitter users seem more resistant to engaging with links to news stories that use fauxtography, but more likely to engage with tweets containing fauxtography images themselves.
Reddit users were more likely to engage with any fauxtography related content.

These differences pose interesting problems for social media platforms; for example, fact-checking efforts that focus on links to news articles (some of which have been implemented by Twitter\edc{cite?}) are likely to have little effect on the spread of fauxtography in general as the images themselves still achieve relatively high engagement.

\section{RQ2: Fauxtography's Evolutionary Nature}

Previous work has indicated that memes exhibit some evolutionary properties, with new variants frequently emerging.
Since, by definition, our fauxtography dataset includes images that have spread wide enough to warrant fact checking, we posit that some might have found life beyond fauxtography.
Thus we ask: do fauxtography images become memes with different variants?

\begin{table*}[t!]
  \centering
  \footnotesize
    \begin{tabular}{lrrrrr}
      \toprule
      \textbf{Platform}  &
       \textbf{\#CommonFalse-} &
       \textbf{\#FalseImages with} &
       \textbf{\#FalseImages w/o} &
       \textbf{\#FalseImages w/o} &
       \textbf{\#FalseImages with}\\

       &
       \textbf{NoRating Images} &
       \textbf{variation} &
       \textbf{any variation} &
       \textbf{variation-RandomImages} &
       \textbf{variation-SameImage}\\

      \midrule
      Twitter &                  {162} &                    {86} &                     {76}
      &                      {1291}
      &                    {70} \\
      Reddit &                 {145} &                   {70} &                    {75}
      & {625}
      &{63} \\
      4chan &                   {58} &                      {25}              &          {33}
      & {269}
      & {117} \\
      \bottomrule
    \end{tabular}%
  \reduce\reduce
  \caption{Statistics for false images variations in Twitter, Reddit, and 4chan.}
  \label{tbl:falseimages}
  \reduce\reduce
\end{table*}

To answer this question, we relax our distance threshold used to detect instances of fauxtography images from 6 to 8 and examine the resulting images matches.
We further focus on fauxtography images that were labeled only False, to focus on the role of fauxtography in spreading false information.
We find 238 source images labeled ``False'' from our Snopes dataset that appear at least once on all Twitter, Reddit, and 4chan.
We note that although measuring engagement on 4chan is problematic enough that we do not include details in Section~\ref{sec:engagement}
, 4chan is a key player in the meme ecosystem; thus we include it in this analysis.

For each source image, we manually determine whether each image within distance 8 is a variant.
\jbnote{More words when the numbers are solidified and explained}\ft{For example out of these 238 false common images between all three platforms, 162 of them on Twitter have images with distance equals to 8. From these 162 images 86 images has at lease one image which is the variation of the False image, and 76 of them have no variations. The last two columns shows the number of False images from each platform that there is no variation of them have been found in the dataset. This can be because of the reason that images with distance 8 from the Snopes image can be totally some random images which is shown in column 5 of the table, or the images with distance 8 were the same Snopes image but being resized or with different quality which is shown in column 6 of the table.}
Table~\ref{tbl:falseimages} provides details on the number of instances of variants across each platform.
We observe that, of the 238 source images appearing on all three platforms, there were an additional 162 images on Twitter within distance 8 that we confirmed were indeed a match for a source image.
Of these 162, 86 were sufficiently different from the source image to be deemed a variant, while 76 were essentially the same as the source image (i.e., they can be considered false negatives due to our threshold selection).
An additional 1,291 images with distance 8 were completely unrelated (i.e., true negatives).
For each of the three platforms, we see relatively similar numbers.
\jbnote{@FATEMEH, check the previous few lines. I am still not sure what the last column is in the table; not sure if we should include it at all; we also need to change the column headers in the table, but that's simple.}

\begin{figure*}[t!]
\centering
\subfigure[Al Franken source image]{
\includegraphics[width=0.45\columnwidth]{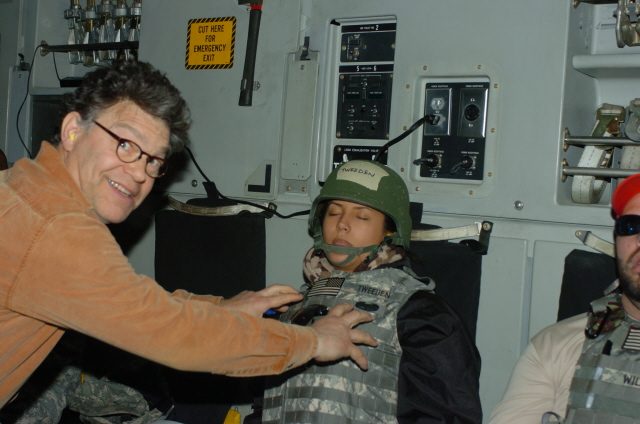}
\label{fig:franken_snopes}
}%
\vspace{-0.05cm}
\subfigure[Al Franken Twitter variant]{
\includegraphics[width=0.45\columnwidth]{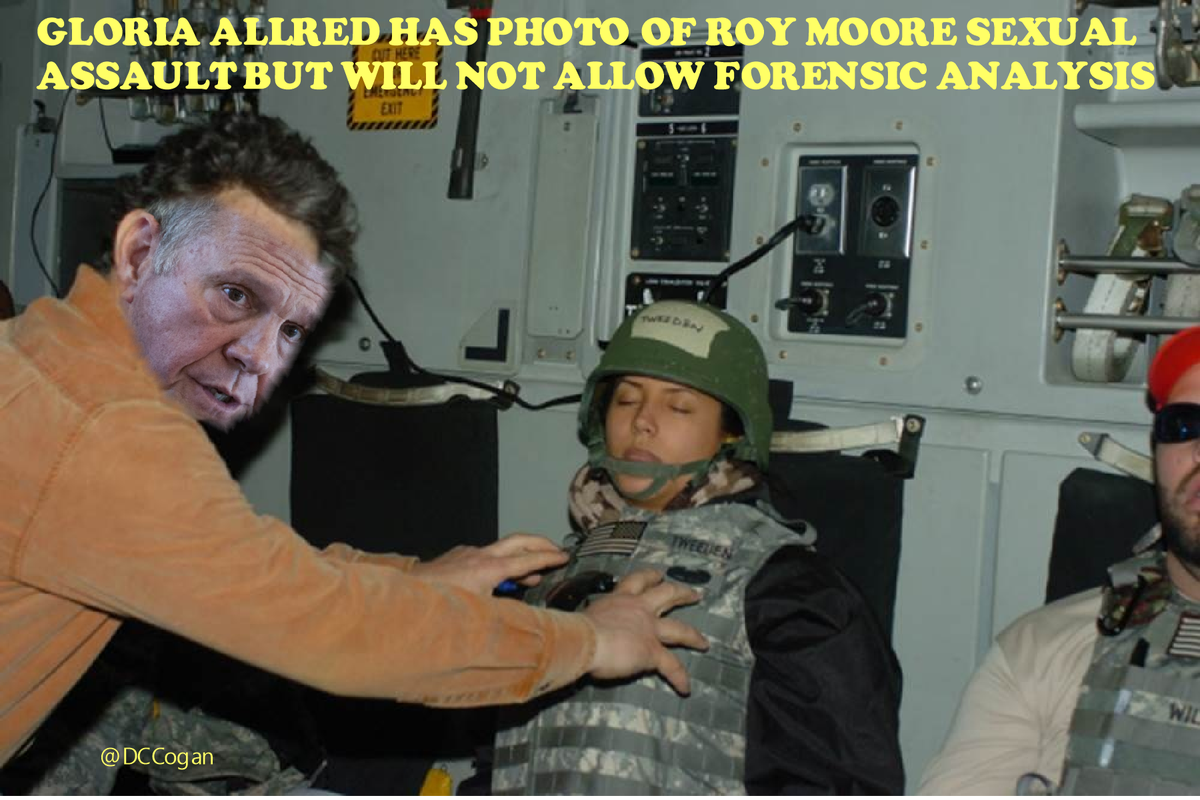}
\label{fig:franken_twitter}
}%
\vspace{-0.05cm}
\subfigure[Al Franken Reddit variant]{
\includegraphics[width=0.45\columnwidth]{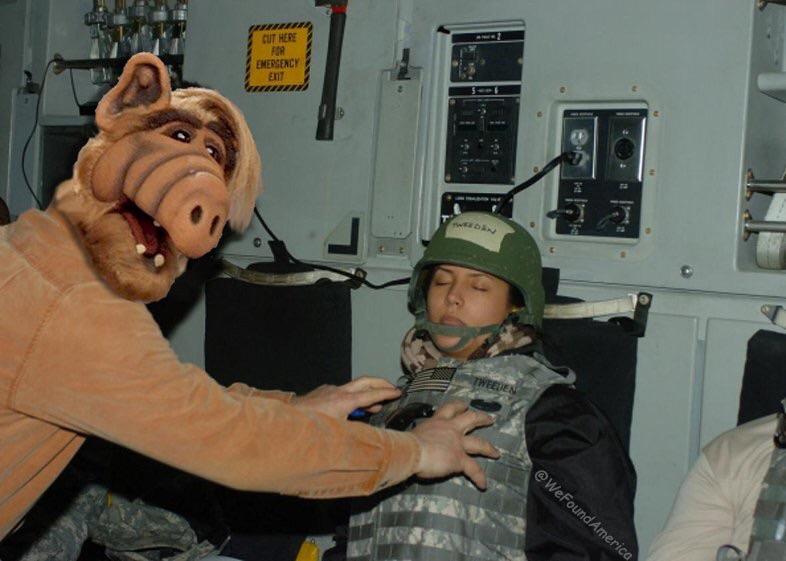}
\label{fig:franken_reddit}
}%
\vspace{-0.05cm}
\subfigure[Al Franken 4chan variant]{
\includegraphics[width=0.45\columnwidth]{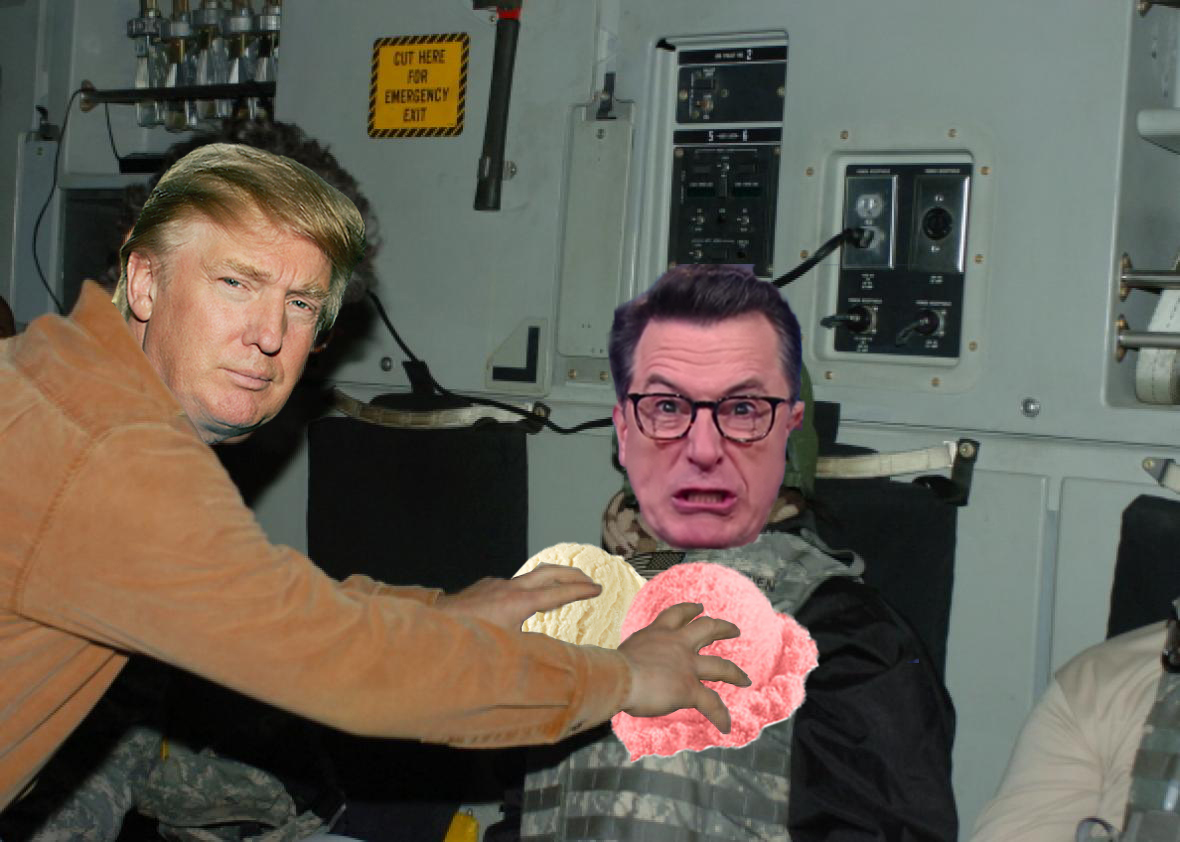}
\label{fig:franken_4chan}
}%
\vspace{-0.05cm}
\newline
\subfigure[Trump Marine 1 original]{
\includegraphics[width=0.45\columnwidth]{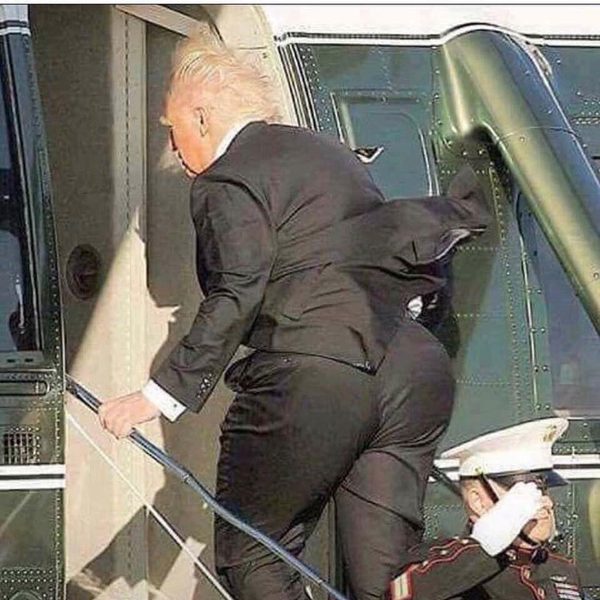}
\label{fig:trump_butt_snopes}
}%
\vspace{-0.05cm}
\subfigure[Marine 1 Twitter variant]{
\includegraphics[width=0.45\columnwidth]{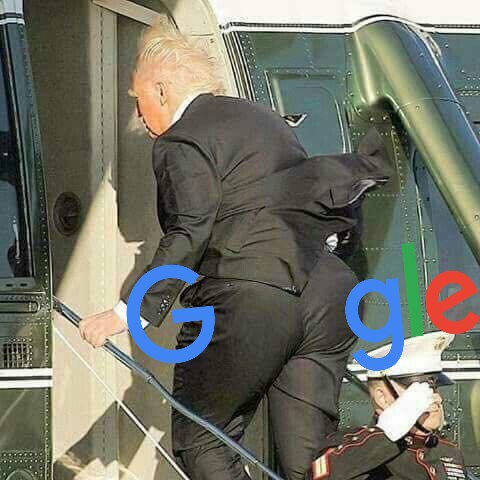}
\label{fig:trump_butt_twitter}
}%
\vspace{-0.05cm}
\subfigure[Marine 1 Reddit variant]{
\includegraphics[width=0.45\columnwidth]{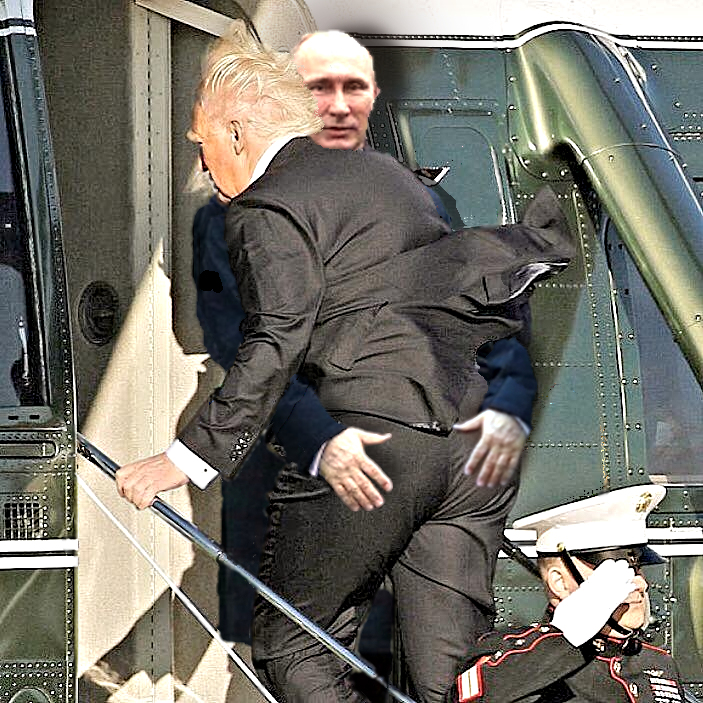}
\label{fig:trump_butt_reddit}
}%
\vspace{-0.05cm}
\subfigure[Marine 1 4chan variant]{
\includegraphics[width=0.45\columnwidth]{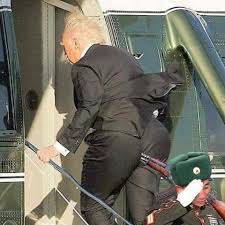}
\label{fig:trump_butt_4chan}
}%
\vspace{-0.05cm}
\newline
\subfigure[GWB reading book original]{
\includegraphics[width=0.45\columnwidth]{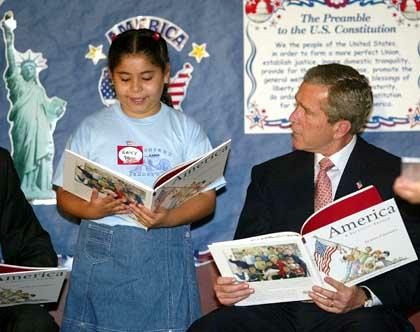}
\label{fig:gwb_book_snopes}
}%
\vspace{-0.05cm}
\subfigure[GWB Twitter variant]{
\includegraphics[width=0.45\columnwidth]{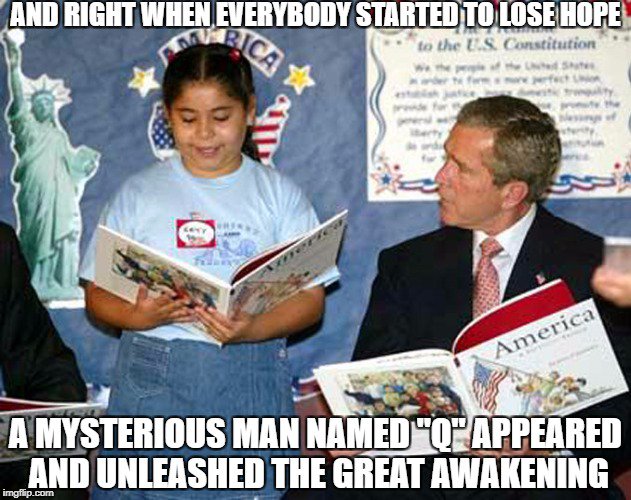}
\label{fig:gwb_book_twitter}
}%
\vspace{-0.05cm}
\subfigure[GWB Reddit variant]{
\includegraphics[width=0.45\columnwidth]{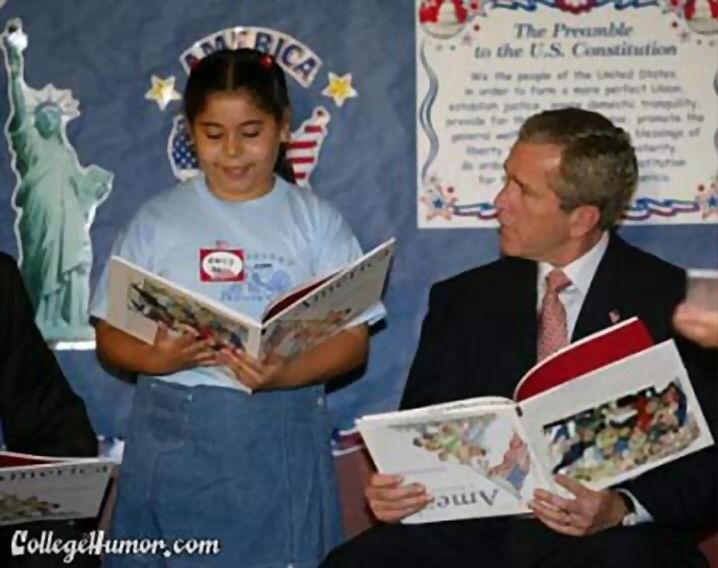}
\label{fig:gwb_book_reddit}
}%
\vspace{-0.05cm}
\subfigure[GWB 4chan variant]{
\includegraphics[width=0.45\columnwidth]{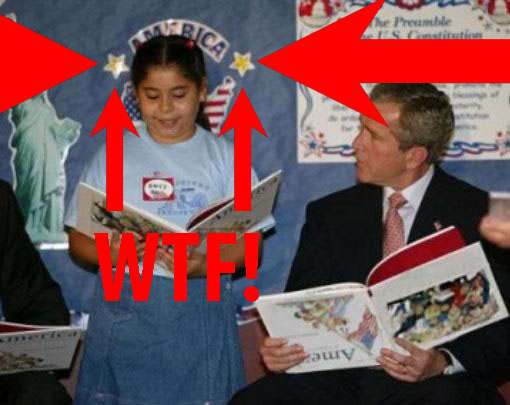}
\label{fig:gwb_book_4chan}
}%
\vspace{-0.15cm}
\caption{Variations of three common False images on all three platforms.}
\vspace{-0.15cm}
\label{fig:variants}
\end{figure*}

\subsection{Case Studies}
\bb{attempt at transitional text here} We find that 13 source images have variants that appear at least once on all three platforms we study (although not necessarily the same variant).
A manual inspection shows that variants of these 13 source images correspond to memes.
We examine three representative and particularly well-known cases in Figure~\ref{fig:variants}.
Our intuition is that particularly powerful fauxtography images are likely to take on a life of their own and become memes.

The first source image (Figure~\ref{fig:franken_snopes}) is a picture of Al Franken inappropriately touching Leeann Tweeden's breasts while she slept.
The image is real, and was taken in 2006 on a C-17 cargo plane on their return from a USO tour in Afghanistan.
This source image played a crucial role in then US Senator Al Franken's retirement from politics.
The image was particularly controversial due it coming to light at the height of the \#MeToo movement~\jbnote{there is some me too paper we can cite too}\cite{metooTheAtlantic} as well as claims that it was related to a sketch that had been performed on the USO tour.
The image is labeled as a false instance of fauxtography due to a widely circulated claim that the photographer that took the picture said it was staged.
However, Franken fully admitted to the picture to be be real and not staged, accepted responsibility for what was ultimately irresponsible behavior, and resigned.

The variant of this image on Twitter (Figure~\ref{fig:franken_twitter}) replaces Franken's face with that of Roy Moore, an Alabama political figure that lost a hotly contested race against Democrat Doug Jones for Jeff Sessions's US Senate seat after he was appointed US Attorney General.
The text added to the image is related to allegations of sexual assault and pedophilia by Roy Moore, and Gloria Aldred's involvement in the incident.
The variation on Reddit (Figure~\ref{fig:franken_reddit}) is much less political, merely replacing Franken's head with that of 80s sitcom character Alf.

On 4chan (Figure~\ref{fig:franken_4chan}, the variant replaces Franken's face with with that of Donald Trump, replaces Leeann Tweeden's head with Stephen Colbert's head, and places two scoops of ice cream over Tweeden's breasts.
This is likely in reference to Stephen Colbert's comments on sexual harassment ~\cite{Hoozer2020AnalysisofStephenColbertLateShowMonologues} and Trump's alleged routine of receiving two scoops of ice cream for desert when everyone else at the table receives only one~\cite{CNNtwoIcecreamScopes} (e.g., Colbert's nickname for Trump, ``Donnie Two Scoops'').

The second source image Figure~\ref{fig:trump_butt_snopes} shows a rear view of Donald Trump entering Marine One.
\jbnote{CHECK DESCRIPTION FROM SNOPES AND WRITE SOME STUFF}
Based on Snopes this image is a ``slightly manipulated'' image (Trump's posterior has been enhanced) originally taken by a Reuters' photographer while president Trump boarded Marine One at Joint Base Andrews in Maryland.
This photo was generally considered unflattering for Trump, as can be seen in the Twitter variant (Figure~\ref{fig:trump_butt_twitter}), which uses Trump's buttocks to replace the two ``Os'' in Google's logo.
This is indicative of some of the derision expressed online towards Trump's physical appearance.
The Reddit variant (Figure~\ref{fig:trump_butt_reddit}) introduces Vladimir Putin embracing Trump by ``grabbing his butt.''
The 4chan variant (Figure~\ref{fig:trump_butt_4chan}) is somewhat different, and replaces the saluting Marine guard with a saluting North Korean \jbnote{verify} soldier with a rifle slung over his shoulder.
The end of the rifle barrel is depicted as being inserted into Trump's buttocks.

The final source image we examine (Figure~\ref{fig:gwb_book_snopes}) shows George W. Bush at a book reading at school in Houston in 2002.
Snopes labels it as false because a manipulated version showing Bush holding the book upside down with a false caption was being spread on the Web.
On Twitter (Figure~\ref{fig:gwb_book_twitter}), we see a variant that has a non-manipulated version of the image, but has added text that implies Bush is telling the student about how right when the world needed it, ``Q'' (from the Qanon conspiracy) appeared to save us all.
The variant that appears on Reddit (Figure~\ref{fig:gwb_book_reddit}) is the manipulated variant where it appears Bush is holding the book upside down.
Finally, we see a variant on 4chan (Figure~\ref{fig:gwb_book_4chan}) that uses the manipulated version with the upside down book, and adds large arrows pointing to the stars behind the students along with the text ``WTF!''
We are not entirely sure what this variant is trying to express, but based on our understanding of 4chan, we suspect it is conspiracy theory related.\bb{Maybe too speculative}

\subsection{Takeaways}

Fauxtography is a complicated issue, in large part due to its visual nature and the Web's propensity for not just spreading visual information, but modifying it.
The Franken image, which is not altered in any way and has a known provenance, is easily exploited for uses completely unrelated to its use in fauxtography.
Similarly, the Bush image shows that even relatively innocuous pictures manipulated in subtle ways can become further manipulated to politicize them.
The Trump image shows how even slight manipulations of real photos can elicit numerous meme variants.

This raises serious concerns about how to mitigate the relatively low-tech problem of fauxtography.
For example, none of the variants we found were particularly convincing in terms of being real photos; the majority were very clearly manipulated, as is common for memes.
What is there to fact check about a fictional TV alien groping a sleeping woman, after all?
However, these variants tend to carry the same fundamental idea as the source image that \emph{was} fact checked, and thus can still cause damage.
Although issues like this warrant future exploration, at minimum, they calls into question the efficacy of fact checking \emph{visual} mis/disinformation.

\section{Related work}
In this section, we review previous work on approaches to study and counter broad disinformation efforts and, more closely, on the use of images in mis/disinformation.

\descr{(Textual) Disinformation.}
A large body of research has studied disinformation on social media, with a specific focus on textual content.
\cite{vosoughi2018spread} show that fake news spread faster than true news on Twitter.
By investigating the discussions on mass shooting events on Twitter,~\cite{starbird2017examining} reveals that alternative news outlets actively propagate alternative narratives, while \cite{wilson2018assembling} study information operations through the lens of the ``Aleppo Boy'' narrative, and show that some news media collaborate to spread alternative narratives.
Also, \cite{zannettou2019let} analyze disinformation campaigns carried out by state-sponsored actors, characterizing their influence on social networks, while
\cite{jiang2020modeling} analyze user comments to characterize the public's (dis)belief towards news items.
\cite{flintham2018falling} survey users consuming news on social network and find that both sources and content play key roles in how they evaluate news veracity.
Aiming to detect and counter disinformation, researchers have often relying on machine learning classification~\cite{shu2020hierarchical,wu2018tracing,shu2017fake,castillo2011information,wang2017liar}.
Also,~\cite{bozarth2020toward} introduce a framework to evaluate the performance of different fake news classification models.
For a comprehensive review of work in this space, please refer to~\cite{kumar2018false}.

\descr{Image-based Disinformation.}
More recently, the research community has begun to look at the interplay between images and disinformation. %
\cite{garimella2020images} collect and analyze disinformation images in India from WhatsApp, while \cite{dewan2017towards} present a pipeline to extract themes and sentiments conveyed in images, and highlight several instances where images were used to share disinformation.
\cite{zannettou2018origins} study image memes, showing that they are often used to spread political and hateful content.
\cite{du2020understanding} also focus on memes containing both images and text and find that a third of them are related to politics, also confirming how memes are shared to spread disinformation as well as conspiracy theories.
Finally,~\cite{zannettou2019characterizing} show that Russian-sponsored trolls actively shared politically charged images on Twitter, and that these also influence other social platforms like Reddit, 4chan, and Gab. %

Prior work has also studied fauxtography, aiming to detect false images.
\cite{zhang2018fauxbuster} build a fauxtography detector called ``FauxBuster'' based on machine learning techniques, while~\cite{bayar2016deep} use deep learning to detect manipulated images.
Similarly,~\cite{zlatkova2019fact} extract various features from images and text, and use machine learning to assess the authenticity of specific claims.
Furthermore, they describe which features are the most effective in verifying the authenticity of the claims.

\descr{Remarks.} To the best of our knowledge, our paper is the first to study the effect that fauxtography images have on user engagement on social media, as well as to measure how these images are discussed and shared on different online services.

\section{Discussion \& Conclusion}

In this paper, we presented a data-driven study of fauxtography on social media.
We found that including fauxtography in social media posts increases user engagement, irrespective of the verisimilitude of the fauxtography image. This highlights the need to take images into account when developing disinformation mitigations.
At the same time, we showed that fauxtography images are often taken out of context and turned into memes, which highlights the challenges faced in automatically identifying image-based disinformation.

Next, we discuss the implications of our findings and highlight some limitations of our study.

\descr{Implications of our findings.}
The fact that sharing fauxtography on social media increases user engagement highlights how image-based disinformation cannot be overlooked, and that any effort to curb the problem should take not only text into account, but also images.
At the same time, we showed that fauxtography images are often used as memes on social media, blurring the line between the intention to mislead and satire.
This opens up a number of problems when moderating fauxtography, since it is challenging to automatically determine the intention with which an image is posted, which is often context specific.
Crucially, our study also highlighted the fact that the verisimilitude of fauxtography images does not have an impact on the engagement that they receive.
This suggests that the ``clickbait'' power of these images is what drives engagement, and raises questions on the effectiveness of mitigations based on fact-checking labels and user warnings.

\descr{Limitations.} Naturally, our study is not without limitations.
First, our image analysis pipeline allows us to identify images that are very similar to fauxtography images, but is unable to verify if the image is used in the misleading setting flagged by Snopes.
For example, we are unable to tell if miscaptioned images are being used in a miscaptioned context.
Similarly, for manipulated photos, it is possible that our analysis pipeline identifies the unmodified picture as a fauxtography one.
\gs{Not sure if we want to say anything else}\bb{here is a suggestions}
This motivates future work combining our analysis pipeline with semantic analysis techniques to study the context in which fauxtography is used.
Third, our identification of news outlets using the top 30K Majestic websites excludes many small local news outlets.
Since we expect that local news outlets have less fastidious fact-checking as compared to larger venues, this suggests our analysis will tend to underestimate the spread of fauxtography on the Web.

Additionally, collecting images at scale from the Web present challenges.
In particular, we found that many images were no longer available when we attempted to download them.
Still, we believe that the scale of our dataset is large enough to allow us to gain a comprehensive view of the use of fauxtography on social networks.

\descr{Acknowledgements} This work was partially supported by the Natural Science Foundation under grant CNS-1942610 and by a BU College of Engineering Dean's Catalyst Award. B.B. acknowledges the support of the National Science Foundation under grant no. DMR-1945058.

\small
\bibliographystyle{abbrv}

\end{document}